\documentclass[preprint]{aastex63}

\usepackage[version=4]{mhchem}

\received{June 11, 2021}
\revised{August 4, 2021}
\accepted{August 11, 2021}

\submitjournal{ApJ}

\shorttitle{Photochemistry of Venus-Like Exoplanets}
\shortauthors{Jordan et al.}

\begin{document}

\title{Photochemistry of Venus-Like Planets Orbiting K- and M-Dwarf Stars}

\correspondingauthor{Sean Jordan}
\email{saj49@cam.ac.uk}

\author[0000-0002-2828-0396]{Sean Jordan}
\affiliation{Institute of Astronomy, University of Cambridge, Madingley Rd, Cambridge CB3 0HA, United Kingdom}

\author[0000-0002-7180-081X]{Paul B. Rimmer}
\affiliation{Department of Earth Sciences, University of Cambridge, Downing St, Cambridge CB2 3EQ, United Kingdom}
\affiliation{Cavendish Laboratory, University of Cambridge, JJ Thomson Ave, Cambridge CB3 0HE, United Kingdom}
\affiliation{MRC Laboratory of Molecular Biology, Francis Crick Ave, Cambridge CB2 0QH, United Kingdom}

\author[0000-0002-8713-1446]{Oliver Shorttle}
\affiliation{Department of Earth Sciences, University of Cambridge, Downing St, Cambridge CB2 3EQ, United Kingdom}
\affiliation{Institute of Astronomy, University of Cambridge, Madingley Rd, Cambridge CB3 0HA, United Kingdom}

\author{Tereza Constantinou}
\affiliation{Institute of Astronomy, University of Cambridge, Madingley Rd, Cambridge CB3 0HA, United Kingdom}

\begin{abstract}

Compared to the diversity seen in exoplanets, Venus is a veritable astrophysical twin of the Earth, however its global cloud layer truncates features in transmission spectroscopy, masking its non-Earth-like nature. Observational indicators that can distinguish an exo-Venus from an exo-Earth must therefore survive above the cloud layer. The above-cloud atmosphere is dominated by photochemistry, which depends on the spectrum of the host star and therefore changes between stellar systems. We explore the systematic changes in photochemistry above the clouds of Venus-like exoplanets orbiting K-Dwarf or M-Dwarf host stars, using a recently validated model of the full Venus atmosphere ($0-115\,{\rm km}$) and stellar spectra from the MUSCLES Treasury survey. \ce{SO2}, \ce{OCS} and \ce{H2S} are key gas species in Venus-like planets that are not present in Earth-like planets, and could therefore act as observational discriminants if their atmospheric abundances are high enough to be detected. We find that \ce{SO2}, \ce{OCS} and \ce{H2S} all survive above the cloud layer when irradiated by the coolest K-Dwarf and all seven M-Dwarfs, whereas these species are heavily photochemically depleted above the clouds of Venus. The production of sulfuric acid molecules that form the cloud layer decreases for decreasing stellar effective temperature. Less steady-state photochemical oxygen and ozone forms with decreasing stellar effective temperature, and the effect of chlorine-catalysed reaction cycles diminish in favour of \ce{HO_x} and \ce{SO_x} catalysed cycles. We conclude that trace sulfur gases will be prime observational indicators of Venus-like exoplanets around M-Dwarf host stars, potentially capable of distinguishing an exo-Venus from an exo-Earth.

\end{abstract}

\section{Introduction} 
\label{sec:intro}

Venus and Earth are astrophysically indistinguishable yet represent the planetary paradigms of surface hostility versus habitability. Both planets have a similar size, mass, and bulk composition, however their atmospheric compositions and resulting climates are highly divergent \citep{ArneyKane2018, Kane2020}. Whilst Earth maintains habitable surface temperatures, Venus has been pushed into a state of runaway greenhouse, with a \ce{CO2}-dominated atmosphere and thick insulating cloud cover, that heats the surface to $735\,{\rm K}$. This divergence may be linked to Venus' shorter period orbit causing it to receive $1.9$ times the insolation flux that the Earth receives. Based on this distinction, \citet{Kane2014} proposed a `Venus zone' interior to the traditional `Habitable zone' as a diagnostic tool for predicting runaway greenhouse on exoplanets. The outer boundary of the Venus zone remains hypothetical and in detail dependent on myriad geological and atmospheric properties of the planet.  Therefore, confirmation of whether a rocky exoplanet is either Earth-like or Venus-like can ultimately only be made by atmospheric characterisation.

Distinguishing between Venus-like exoplanets and Earth-like exoplanets requires observational discriminants in transmission or emission spectroscopy, however the high altitude cloud deck of a Venus-like exoplanet will truncate spectral features from the atmosphere below the clouds. \citet{Lustig-Yaeger2019} show that observations of Venus-like exoplanets in transmission spectroscopy are sensitive to the altitude of the cloud top and observations in emission spectroscopy are more sensitive to the total optical depth of the clouds in the viewing column. Emission spectroscopy of Venus-like exoplanets within an optimally positioned transparency window at $6\,{\rm \mu m}$ can probe higher pressures than in transmission spectroscopy, however emission observations in transparency windows still cannot probe within an order of magnitude of surface pressure, leaving the lower atmosphere out of reach \citep{Lustig-Yaeger2019}. \citet{Wildt1940} used the saturated emission observations of \ce{CO2}, originating from above the clouds of Venus, to predict that the surface temperature would likely be higher than the boiling point of water. However \citet{BennekeSeager2012} show that for exoplanets with cloudy \ce{CO2}-\ce{N2} atmospheres, spectral features for the \ce{CO2}-dominated case were similar to those of the \ce{N2} dominated case, because increases in \ce{CO2} are compensated for by decreased atmospheric scale height. This effect is in addition to potential photochemical clouds and hazes in \ce{CO2}-dominated Venus-like atmospheres, which further flatten spectral features and mask the Rayleigh scattering slope. \citet{Ehrenreich2012} and \citet{MunozMills2012} show the effect of photochemical hazes flattening spectral features in synthetic spectra of Venus' atmosphere, with Mie scattering from upper atmosphere haze dominating the spectra. \citet{Barstow2016} compared synthetic spectra of Earth and Venus as transiting exoplanets, and found that an Earth-like planet's spectrum is unambiguously characterised by a strong Ozone feature whereas a Venus-like planet's spectrum is flat, containing strong features only from \ce{CO2} absorption bands that are also found for the Earth-like case, and cannot be conclusively assigned to either the Earth or Venus atmosphere. Correctly distinguishing Venus-like from Earth-like atmospheres by their spectra requires identifying spectrally active gases present on Venus and that are not present in Earth-like atmospheres. Fulfilling this role is ozone on Earth, and potentially \ce{SO2} and \ce{OCS} on Venus, if their abundances above the cloud layer were high enough to be detected. However, relying on ozone as an observational discriminant is complicated by the discovery of an ozone layer on Venus \citep{Montmessin2011, Marcq2019}, and exacerbated by the potential for massive abiotic \ce{O2} build up on exoplanets around M-Dwarf host stars \citep{Gao2015, LugerBarnes2015}.

Ruling out ozone places the emphasis on positively identifying Venus-like atmospheric constituents. The challenge with this is the photochemical clouds created on Venus-like planets, which lead to only the tenuous region above the cloud layer being observationally accessible.  This above-cloud region of the atmosphere is dominated by photochemistry \citep{YungDeMore1982}, which depends on the spectrum of the host star. The detection bias toward finding exoplanets around smaller stars has led to the discovery of many transiting exoplanets around K-Dwarfs and M-Dwarfs that are ideal for future atmospheric characterisation. The additional detection bias of identifying planets on shorter period orbits means that these planets preferentially reside within the hypothetical Venus Zone, rather than the Habitable Zone, of their host stars \citep{Kane2014}. 

The above-cloud atmospheres of Venus-like exoplanets around K- and M-type host stars will be intrinsically different to the atmosphere of Venus due to the different irradiating starlight. M-Dwarfs can also be highly active, with flaring events and coronal mass ejections occurring more frequently than for Sun-like stars \citep{Gunther2020}. \citet{SchaeferFegley2011} note that such a variable UV input into a planet's atmosphere due to an active M-Dwarf host star could prevent steady-state photochemical cycles from being established. However, even before stellar activity is considered, the different spectra of K-Dwarf and M-Dwarf host stars compared to the Solar spectrum will lead to intrinsic changes in the background atmospheres of their planets, upon which transient perturbations may be introduced by stellar activity. The differences between the stellar spectra of K- and M-Dwarf stars compared to that of the Sun therefore result in changes to the photochemical depletion of trace species, the formation rate of photochemical clouds and hazes, and the rate of catalytic chemistry due to photochemical products, in the atmospheres of these potential Venus-like exoplanets compared to the Venus atmosphere \citep{RugheimerKaltenegger2018, Kane2019}.

We therefore require unique observational discriminants of Venus-like exoplanets around cool stars that are robust to photochemical depletion above the cloud layer by the irradiating starlight. The purpose of this paper is to simulate how the atmosphere of Venus changes when irradiated with a selection of K-Dwarf and M-Dwarf spectra, using a recent atmospheric model validated to Venus observations\footnote{where the model validation requires the addition of a hypothesised scheme of in-droplet chemistry occurring within the cloud droplets \citep{Rimmer2021}, discussed further in Section \ref{sec:model}.}, with the aim of identifying a set of observable chemical species that exist above the cloud layer that identify the planet as Venus-like as opposed to Earth-like. We identify a set of trace sulfur gases that are robust to photochemical depletion of all the M-Dwarf spectra, and find that the rate of cloud formation and catalytic chemistry decreases in correlation with decreasing stellar effective temperature. 

At least one other atmospheric model of Venus, validated by \citet{Lincowski2018}, has been applied to investigate Venus-like exoplanets. The model from \citet{Lincowski2018} couples photochemistry in the Venus atmosphere with climate, cloud, and radiative-transfer modelling in order to obtain synthetic spectra of a range of Venus-like exoplanets around either TRAPPIST-1 \citep{Lustig-Yaeger2019} or Proxima Centauri \citep{Meadows2018a}. Their photochemical model is validated to Venus observations of \ce{H2O}, \ce{SO}, \ce{SO2}, and \ce{OCS}, extends from $28$ -- $112\,{\rm km}$ altitude, and requires surface mixing ratios of \ce{SO2} lower than the observed values in order to reproduce \ce{SO2} and \ce{H2O} observations above the clouds. Other studies have investigated synthetic spectra of Venus-like exoplanets in transmission or emission spectroscopy using prescribed atmospheric mixing ratios rather than coupled photochemistry, and provide insight into the resulting spectral features as well as the inherent ambiguity in retrievals due to the Venusian cloud deck \citep[e.g.,][]{Barstow2016}. Other models of Venus exist which couple thermochemistry and photochemistry for the full atmosphere ($0 - 115\,{\rm km}$)  \citep[e.g.,][]{BiersonZhang2020} but have not been tested with irradiating stellar fluxes other than the Sun.

This study complements past work by being the first to use a detailed and well validated chemical model of the full atmosphere of Venus in the investigation of Venus-like exoplanets. The reason this is possible is because we do not couple climate modelling or generate synthetic spectra, but rather focus on the systematic effects on the atmospheric chemistry that result from irradiation by K-Dwarf and M-Dwarf host stars. Where past models have used a single host star and simulate a broader range of possible atmospheric compositions and climates of Venus-like exoplanets, this study utilises the whole set of stellar spectra from the MUSCLES Treasury survey, and restricts the atmosphere to close Venus-analogues. This enables us to identify general trends in the planetary atmospheres with stellar effective temperature, where all other atmospheric inputs remain unchanged.

The paper is organised as follows. Section \ref{sec:model} describes the modelling methods used for the reference model of Venus in the Solar System, and Venus-like exoplanets in different stellar systems. Section \ref{sec:results} presents our results, first for the photochemical parent species in the atmosphere, and then for the photochemical product species that form as a result of the incident starlight. We discuss the significance of our results and assumptions in Section \ref{sec:discussion} and end with our conclusions in Section \ref{sec:conclusion}.

\section{The model} 
\label{sec:model}

This study uses the photochemical-diffusion model from \citet{Rimmer2021}. It is based off the model from \citet{RimmerHelling2016} and \citet{RimmerRugheimer2019} and has been validated to the atmosphere of Venus to within an order of magnitude of most observations. The model is composed of a $1{\rm D}$ Lagrangian solver, ARGO, and a network of reactions, STAND2020. STAND2020 is a chemical network that treats H/C/N/O/S/Cl chemistry accurately within a temperature range of $100 - 30,000\,{\rm K}$ \citep{RimmerHelling2016,Hobbs2020}. The network contains over $3000$ chemical reactions to be solved by ARGO, as well as condensation chemistry for \ce{H2SO4} and sulfur allotropes. Other elements are included but we do not focus on them in this study. At each altitude step in the atmosphere the reactions are solved by ARGO as a set of time-dependent, coupled, non-linear differential equations:

\begin{equation}
\dfrac{dn_{\rm X}}{dt} = P_{\rm X} - L_{\rm X}n_{\rm X} - \dfrac{\partial \Phi_{\rm X}}{\partial z},
\end{equation}
where, at a given height $z\,{\rm (cm)}$ and time $t\,{\rm (s)}$, $n_{\ce{X}}$ (cm$^{-3}$) is the number density of species $\ce{X}$, $P_{\rm X}$ (cm$^{-3}$ s$^{-1}$) is the rate of production of species \ce{X}, $L_{\rm X}$ (s$^{-1}$) is the rate constant for destruction of species \ce{X}, and $\partial \Phi_{\rm X}/\partial z$ (cm$^{-3}$ s$^{-1}$) describes the divergence of the vertical diffusion flux.

\begin{figure*}[ht!]
\gridline{\fig{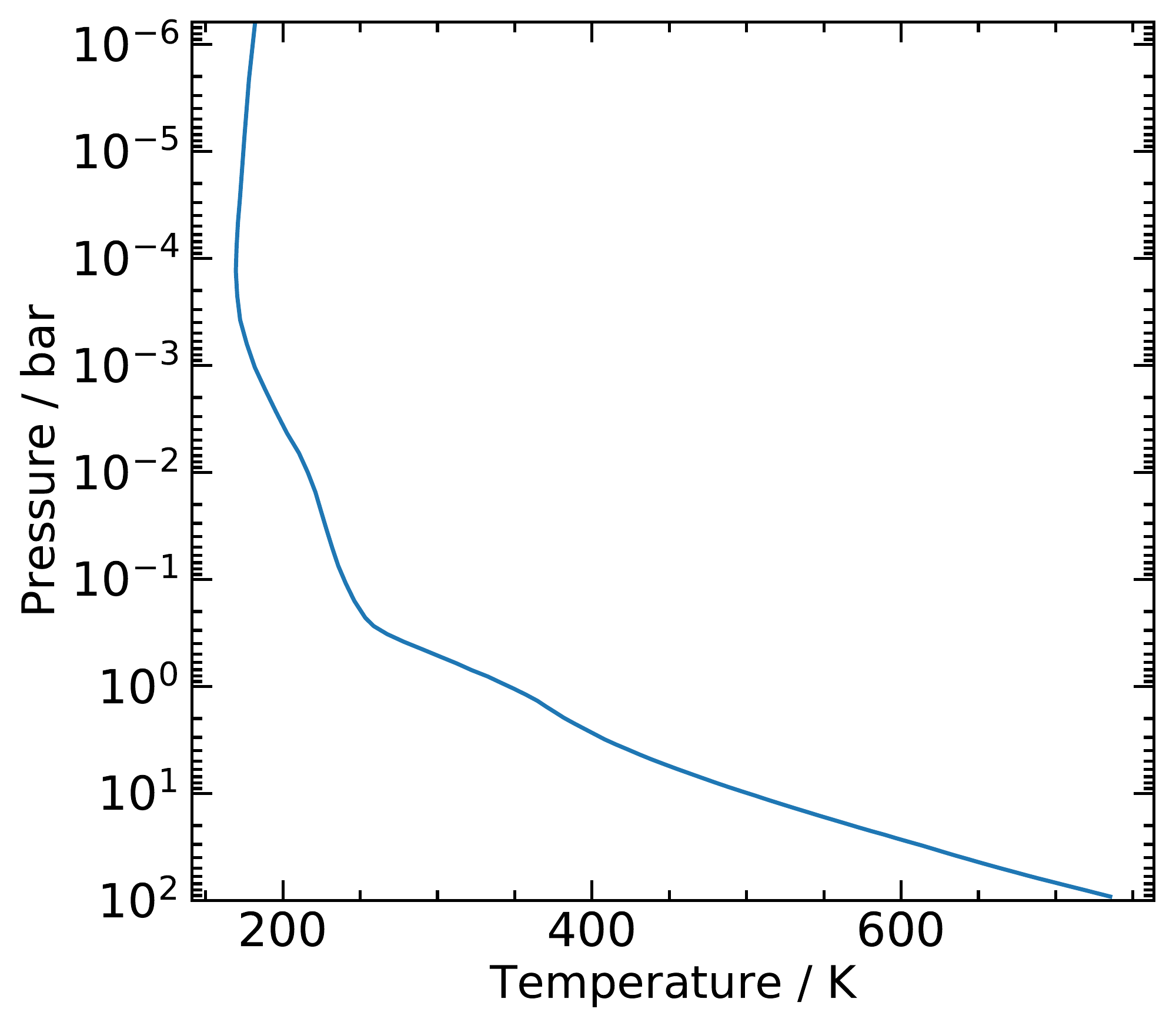}{0.45\textwidth}{}
          \fig{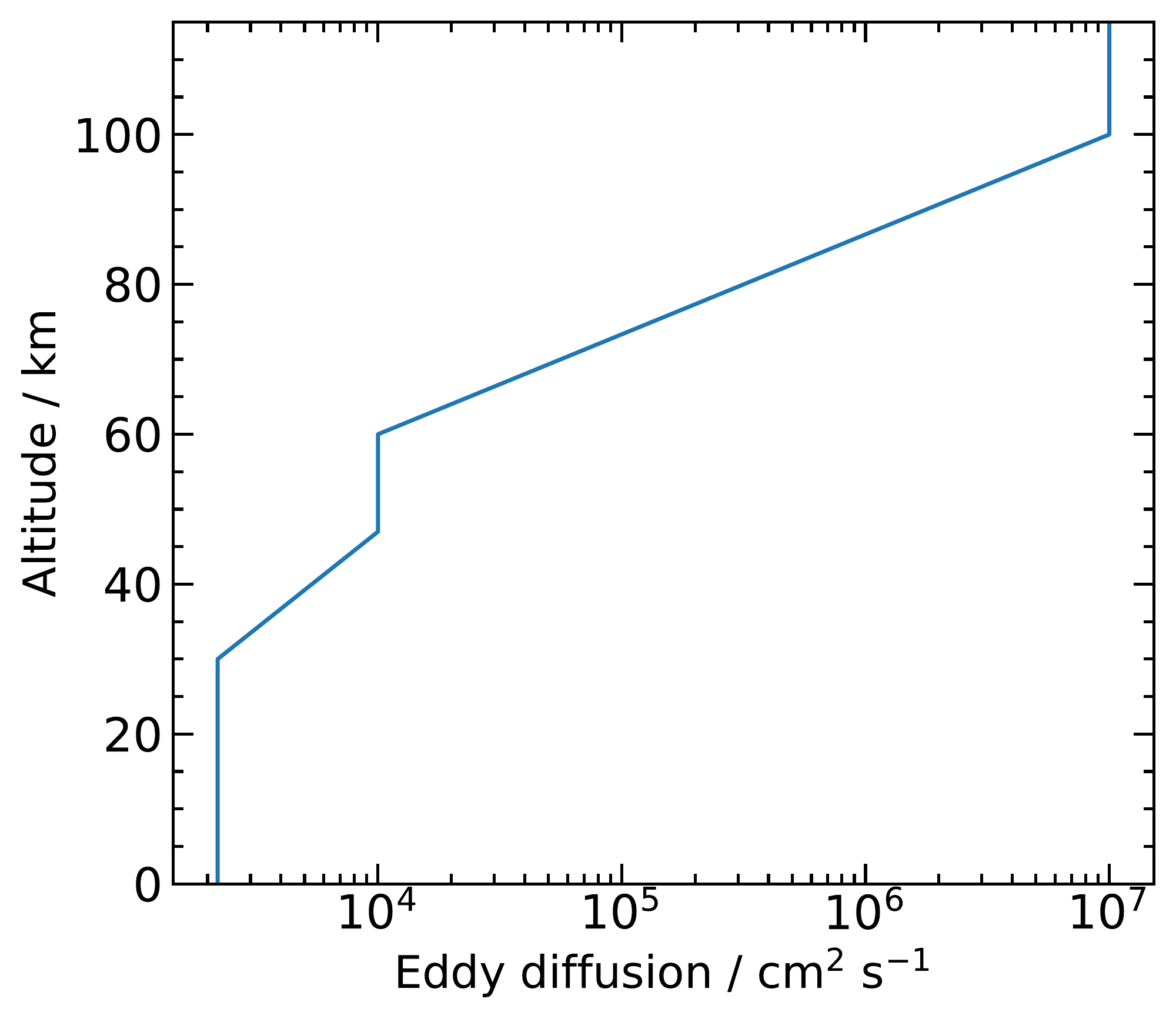}{0.45\textwidth}{}
          }
\caption{Pressure-temperature profile (\textit{left}) \citep{Kras2007}, and eddy diffusion profile (\textit{right}) (\citet{Kras2007}; \citet{Kras2012}) model inputs. \label{fig:pt_eddy}}
\end{figure*}

ARGO follows a parcel of gas as it rises from the surface to the top of the atmosphere and back down again. An initial condition for the chemical composition is input at the base of the atmosphere, and at every altitude step on the journey upwards ARGO calculates thermochemical equilibrium for gas phase chemistry determined by pressure and temperature. The pressure-temperature profile is shown in Figure \ref{fig:pt_eddy} \citep{Kras2007}, and Table \ref{tab:init} lists the initial chemical composition input at the base of the atmosphere. \citet{Rimmer2021} show that the values in Table \ref{tab:init} are consistent with the interpretation that the near-crust atmosphere and solid surface rock are in thermochemical- and phase-equilibrium (except for \ce{NO} whose chemical origin is expected to be via lightning processes). At the top of the atmosphere, ARGO takes the incident stellar spectrum and at each step on the journey downwards, solves for both thermochemical equilibrium and the photochemical reactions driven by the stellar flux irradiating that layer of the atmosphere. The time interval over which the chemistry is solved for at each altitude step depends on the timescale for vertical transport, determined by the eddy diffusion profile shown in Figure \ref{fig:pt_eddy} \citep{Kras2007,Kras2012}.

\citet{Rimmer2021} validate the model against an exhaustive set of observations of the Venus atmosphere that have been collated from the literature. These observations were made at a multitude of different times and of spatial locations within the Venusian atmosphere and as a result the data can often conflict with each other when expressed in a 1D model. This poses a challenge in constructing detailed 1D atmospheric models in general. Nonetheless, the model we employ is able to match most atmospheric data on Venus from the surface to $115\,{\rm km}$ to within 1 order of magnitude, making it ideally suited to generalise to Venus-like planets around other stars.

In order to successfully reproduce observations of \ce{SO2} and \ce{H2O} in the middle atmosphere, \citet{Rimmer2021} hypothesise a scheme of aqueous droplet chemistry occurring within the clouds. The hypothesis is motivated by the fact that the abundances of \ce{SO2} and \ce{H2O} below the clouds are not consistent with the previously held assumption that \ce{SO2} depletion in the middle atmosphere is due entirely to the \ce{H2SO4} formation reaction. There is currently no whole-atmosphere model of Venus that can predict the \ce{SO2} and \ce{H2O} depletion and this is a known and long-standing problem \citep{BiersonZhang2020, Marcq2018}. The proposed droplet chemistry remedies this by enhancing the \ce{SO2} depletion that occurs through the cloud-forming region of the atmosphere by including aqueous chemistry for dissolved \ce{SO2} within the droplets. The chemical scheme requires the presence of mineral salt dissolved in the \ce{H2SO4 (l)} droplets which is assumed to originate from a combination of vertical transport of surface dust to the clouds, volcanism, and exogeneous delivery. The rate of reaction for this droplet chemistry is unconstrained, depending on the mass and composition of the mineral dust, and has been tuned to match observations of Venus. 

The condensation chemistry for \ce{H2SO4} molecules works via converting \ce{H2SO4 (g)} into \ce{H2SO4 (l)} when the partial pressure of gaseous \ce{H2SO4 (g)} is in excess of the saturation vapour pressure. This prescription does not account for the microphysics of aerosol nucleation, which are beyond the capabilities of the model. The chemical scheme for the proposed droplet chemistry is then modelled by coupling the \ce{H2O (g)} and \ce{SO2 (g)} mixing ratios to the \ce{H2SO4 (l)} mixing ratio via Henry's Law to simulate the dissolution of \ce{H2O (aq)} and \ce{SO2 (aq)} in the cloud droplets. The reation network then includes effective reactions which enable the formation of \ce{H2SO3 (l)} from \ce{H2O (aq)} and \ce{SO2 (aq)}, that are dissolved in the condensed \ce{H2SO4 (l)}. When droplet chemistry is not included in the model, \ce{H2SO4} formation and condensation still occurs, however it then is an insufficient mechanism to produce the observed depletion of \ce{SO2} and \ce{H2O} in the Venusian upper atmosphere.

Without the inclusion of this droplet chemistry, the \ce{SO2} depletion through the cloud layer can otherwise be explained by decreasing the below-cloud \ce{SO2} abundance below the observed values, or by increasing the below-cloud \ce{H2O} abundance and simultaneously introducing a layer of reduced eddy diffusion within the clouds, to inhibit vertical transport and enable the cloud forming reaction to proceed further. Following these alternative hypotheses leads to other species, such as \ce{CO} and \ce{O2}, then not well-fitting the observations, compared to the the case with droplet chemistry included. 

These challenges in accurately modelling Venus' atmospheric chemistry in the solar system, create uncertainties in extending the model to Venus-like planets around other stars. For a Venus-like exoplanet, if droplet chemistry is operating then it would not necessarily involve the same salts and would likely proceed at a different rate. We present results in Section \ref{sec:results} for the model with droplet chemistry as this most closely resembles the observed atmosphere of Venus, but we include all results reproduced without droplet chemistry in an appendix. We discuss the extent to which our conclusions for Venus-like exoplanets depend on the inclusion of droplet chemistry in Section \ref{sec:discussion}. Uncertainties around the droplet chemistry hypothesis motivate further investigation and observation of Venus.

\begin{deluxetable*}{cccccccccc}
\tablecaption{Initial Surface Abundances \citep{Rimmer2021} \label{tab:init}}
\tablehead{
\colhead{\ce{CO_2}} & \colhead{\ce{N_2}} & 
\colhead{\ce{SO_2}} & \colhead{\ce{H_2O}} & 
\colhead{\ce{CO}} & \colhead{\ce{OCS}} & \colhead{\ce{HCl}} & \colhead{\ce{H_2}} & \colhead{\ce{H_2S}} & \colhead{\ce{NO}}
} 
\startdata
0.96 & 0.03 & 150 ppm & 30 ppm & 20 ppm & 5 ppm & 500 ppb & 3 ppb & 10 ppb & 5.5 ppb\\
\enddata
\end{deluxetable*}

In order to apply the model to Venus-analogue exoplanets, we keep all inputs the same except for the stellar flux incident at the top of the atmosphere. The set of stellar spectra we use are from the MUSCLES Treasury Survey \citep{France2016, Youngblood2016, Loyd2016}, and correspond to four K-Dwarf and seven M-Dwarf stellar energy distributions. Table \ref{tab:muscles} lists their stellar parameters. Since K-Dwarf and M-Dwarf stars are intrinsically lower luminosity than the Sun, the flux profiles must each be scaled to a Venus-equivalent flux, corresponding to the orbital distance at which an exoplanet would receive the same total incident flux of starlight as Venus receives from the Sun. This scaling was performed by using the ratio of the bolometric flux of each spectrum, as calculated by the MUSCLES team, to the bolometric flux at Venus' orbital distance from the Sun. 

Scaling to a Venus-equivalent flux is a first approximation to self-consistency in the temperature profile of the model atmospheres. The atmospheric composition, and the radiative effects of greenhouse gases, will vary with changes in the stellar spectrum. This will alter the temperature profile of the atmosphere which will in turn alter the altitude at which the temperatures permit condensation and evaporation of \ce{H2SO4} cloud droplets. Thus, the altitude of the cloud base and cloud top can also vary with changes in the stellar spectrum. Simultaneously calculating a self-consistent temperature profile as the chemistry of the atmosphere changes is beyond the scope of this study and so keeping the bolometric flux incident on the atmosphere constant while assuming a fixed temperature profile is an approximation. This approximation is reasonable for drawing conclusions about the above-cloud region of the atmosphere because this region is dominated by photochemistry rather than thermochemistry. However changes induced in the cloud layer would likely have a significant impact on the temperature profile below the clouds, leading to climate feedbacks that could alter the thermochemical equilibrium in the lower atmosphere and the transport of species between the lower and middle atmospheric regions. We discuss this limitation further in Section \ref{sec:discussion}, and assume that the above-cloud atmosphere, accessible to remote observation, remains the region of the atmosphere above $70\,{\rm km}$ altitude.

\begin{figure*}[ht!]
\gridline{\fig{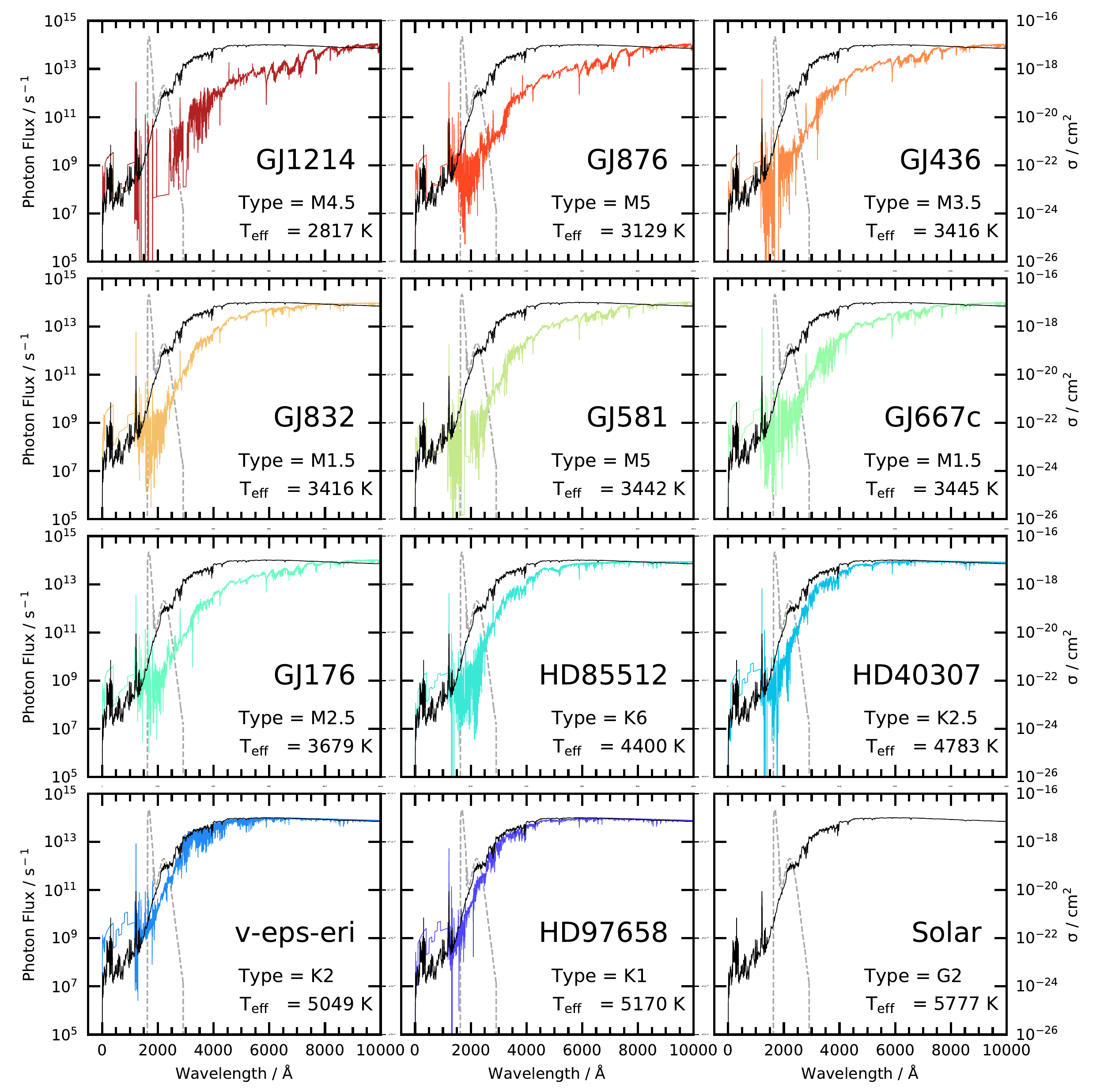}{\textwidth}{}
         }
\caption{Stellar spectra of the MUSCLES stars compared to the Solar spectrum (black), in order of increasing literature value of stellar effective temperature. The \ce{OCS} photodissociation cross section, ${\rm \sigma\,[cm^2]}$, is plotted behind (dotted grey) exemplifying how changes in incident UV flux has the potential to alter the atmospheric composition via photochemistry. This is particularly prevelant for \ce{OCS} in our exo-Venus model atmospheres (see section \ref{sec:phot_parent}.) \label{fig:muscles_spectra}}
\end{figure*}

The set of scaled stellar spectra are shown in Figure \ref{fig:muscles_spectra}. They are presented in order of increasing stellar effective temperature with a colour map running from red (coolest) to blue (hottest). The Solar spectrum at Venus-distance is shown alongside for reference. As the stellar effective temperature increases the peak of the flux distribution shifts to shorter wavelengths, eventually converging with the Solar flux distribution longward of $\sim 2000\,{\rm \AA}$. Shortward of $\sim 2000\,{\rm \AA}$, in the Far Ultra-Violet region (FUV), there is considerable spread in the flux distribution down to the strong Lyman-alpha line at $\sim 1200\,{\rm \AA}$. Below $\sim 1200\,{\rm \AA}$, in the Extreme Ultra-Violet region (EUV), the flux is similar to that of the Sun for the cool M-Dwarfs and increases with stellar effective temperature until it is generally greater than Solar at most EUV wavelengths for the four K-Dwarfs. The different UV inputs into the model atmospheres leads to changes in the photochemical depletion of different species depending on the wavelength distribution of their photodissociation cross sections. As an example of this, alongside the stellar spectra in figure \ref{fig:muscles_spectra} we plot the photodissociation cross section of \ce{OCS}. 

\begin{deluxetable*}{cccccc}
\tablecaption{Stellar parameters of the MUSCLES stars\label{tab:muscles}}
\tablehead{
\colhead{Star} & \colhead{\sc{MUSCLES} T$_{\rm{eff}}$} [K] & 
\colhead{Literature T$_{\rm{eff}}$} [K] &
\colhead{Distance [pc]} & \colhead{Radius [R$_\odot$]} & 
\colhead{Spectral Type}
} 
\startdata
GJ1214 & 2935 & 2817 & 14.6 & 0.21 & M4.5 \\
GJ876 & 3062 & 3129 & 4.7 & 0.38 & M5 \\
GJ436 & 3281 & 3416 & 10.1 & 0.45 & M3.5 \\
GJ581 & 3295 & 3442 & 6.2 & 0.30 & M5 \\
GJ667c & 3327 & 3445 & 6.8 & 0.46 & M1.5 \\
GJ176 & 3416 & 3679 & 9.3 & 0.45 & M2.5 \\
GJ832 & 3816 &  3416 & 5.0 & 0.56 & M1.5 \\
HD85512 & 4305 & 4400 & 11.2 & 0.70 & K6 \\
HD40307 & 4783 & 4783 & 13.0 & 0.83 & K2.5 \\
HD97658 & 5156 & 5170 & 21.1 & 0.72 & K1 \\
v-eps-eri & 5162 & 5049 & 3.2 & 0.74 & K2.0 \\
\enddata
\end{deluxetable*}

\section{Results} 
\label{sec:results}

The species listed in Table \ref{tab:init} are the photochemical parent species from which the rest of the atmospheric composition follows. Their abundances are affected primarily by thermochemistry below the cloud layer and by photochemistry above. We present their mixing ratios as a function of altitude and pressure in Section \ref{sec:phot_parent}. We then examine the photochemical product species that form as a result of this photochemistry in Section \ref{sec:sulfur_species} and \ref{sec:oxygen_cycles}. These photochemical product species include \ce{SO}, \ce{SO3}, and \ce{H2SO4 (l)}, resulting from \ce{SO2} photochemistry, and \ce{O2} and \ce{O3}, resulting from catalytic formation from \ce{O} atoms.

\subsection{Photochemical parent species} 
\label{sec:phot_parent}

\begin{figure*}[ht!]
\gridline{\fig{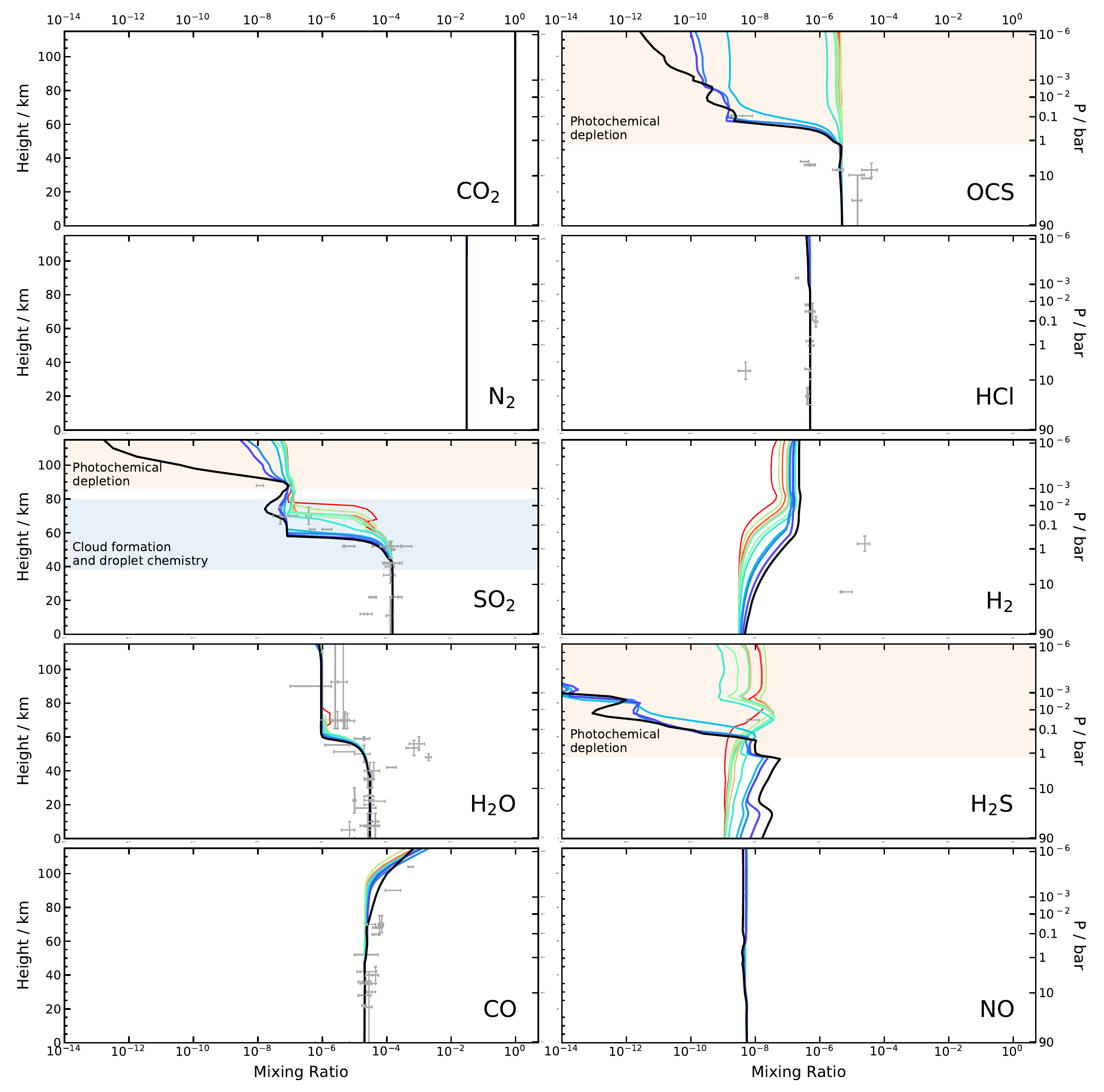}{\textwidth}{}}
\gridline{\fig{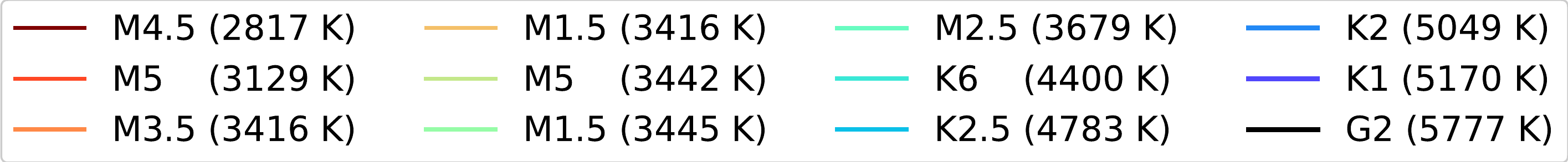}{0.8\textwidth}{}}
\caption{Photochemical parent species in the atmosphere, in order of decreasing surface mixing ratio. Droplet chemistry is included.  \label{fig:phot_parents_cloud_chem}}
\end{figure*}

Figure \ref{fig:phot_parents_cloud_chem} shows the behaviour of the photochemical parent species for all twelve host stars, with a colourmap from the lowest effective temperature M-Dwarf in red to the highest effective temperature K-Dwarf in blue. The behaviour of these species on Venus is shown for reference in black.

The trace sulfur gases \ce{SO2}, \ce{OCS}, and \ce{H2S} are heavily photochemically depleted on Venus by the Solar flux, however all survive above the cloud layer for Venus-like exoplanets around the coolest K-Dwarf, HD85512, and all seven M-Dwarf host stars. On Venus, the \ce{SO2} mixing ratio decreases between $45$ -- $60\,{\rm km}$ altitude due to cloud formation and is then photochemically depleted by over five orders of magnitude above $90\,{\rm km}$ by Solar UV. Meanwhile above $50\,{\rm km}$, \ce{OCS} is photochemically depleted by six orders of magnitude and \ce{H2S} is entirely photochemically depleted by the Solar UV. The depletion of each of these species decreases as stellar effective temperature decreases until, for all seven M-Dwarf host stars, there is no longer any photochemical depletion and the three species survive with much greater above-cloud abundance than on Venus.

Whilst the remaining species in Figure \ref{fig:phot_parents_cloud_chem} undergo photochemistry above the cloud layer, the rate of their photochemical reactions is not sufficient to deplete them on Venus, and this remains true for Venus-like exoplanets around the M-Dwarf and K-Dwarf host stars. \ce{CO2}, \ce{N2}, \ce{HCl} and \ce{NO} all have constant mixing ratios through the atmosphere of Venus, and no changes are incurred by the different stellar spectra. \ce{H2O}, \ce{CO}, and \ce{H2} do not have constant mixing ratios on Venus but undergo only minor changes due to the different stellar spectra. \ce{SO2}, \ce{OCS} and \ce{H2S} are the only photochemical parent species that vary significantly, with above-cloud abundances many orders of magnitude greater than their abundances on Venus. In Section \ref{sec:discussion} we discuss how this may influence the observable spectral features that we can expect from Venus-like exoplanets around M-Dwarfs and K-Dwarfs. 

\subsection{Sulfur species production} 
\label{sec:sulfur_species}

The different rate of \ce{SO2} photochemistry has a knock on effect to photochemical product species in the atmosphere. The main photochemical reaction of \ce{SO2} produces \ce{SO} and \ce{O} above $\sim$ 60 km in the atmosphere by
\begin{align}
\ce{SO2}  + h\nu &\rightarrow \ce{SO} + \ce{O}.
\label{eq:SO2_hv}
\end{align}

\begin{figure*}[ht!]
\gridline{\fig{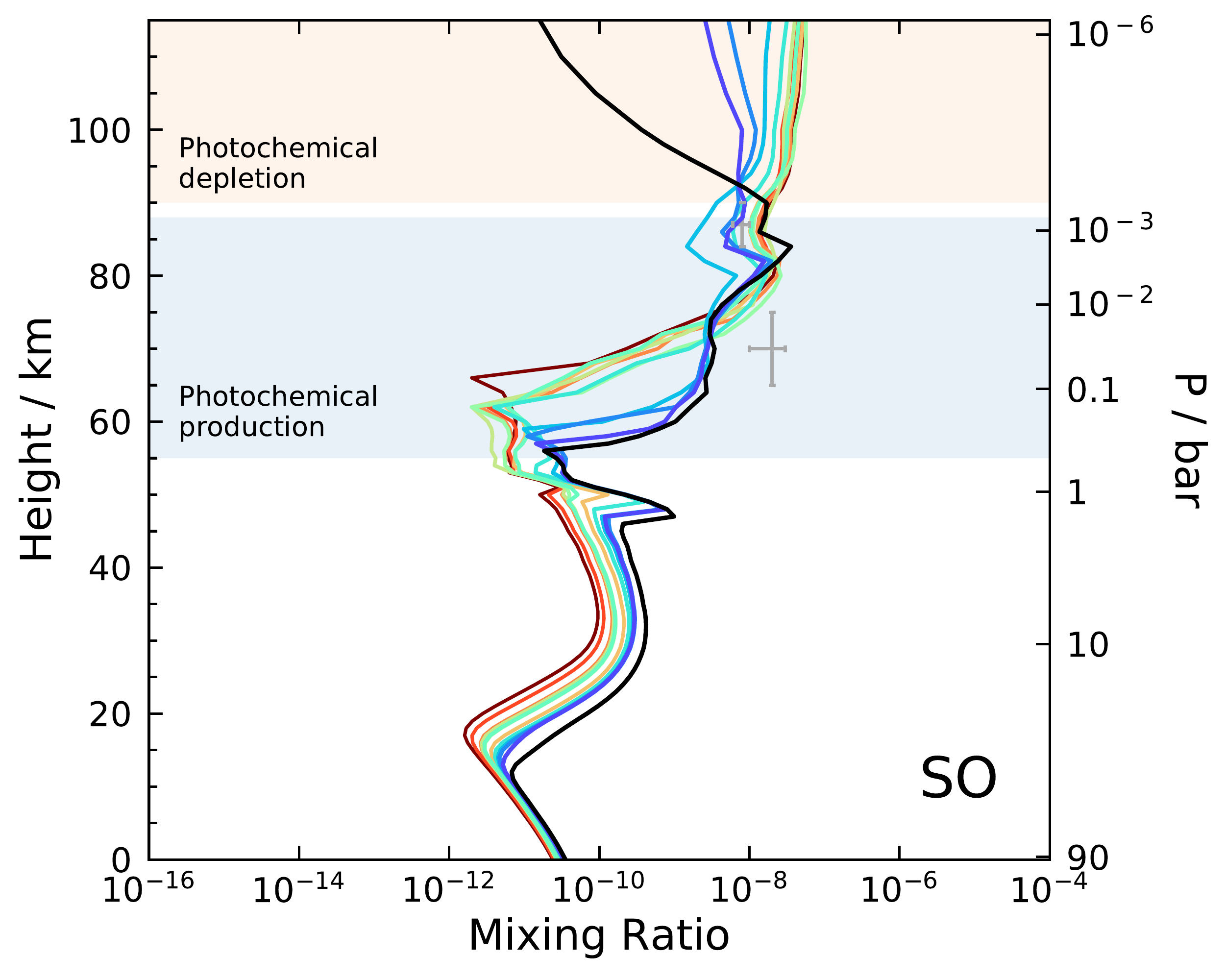}{0.45\textwidth}{}
          \fig{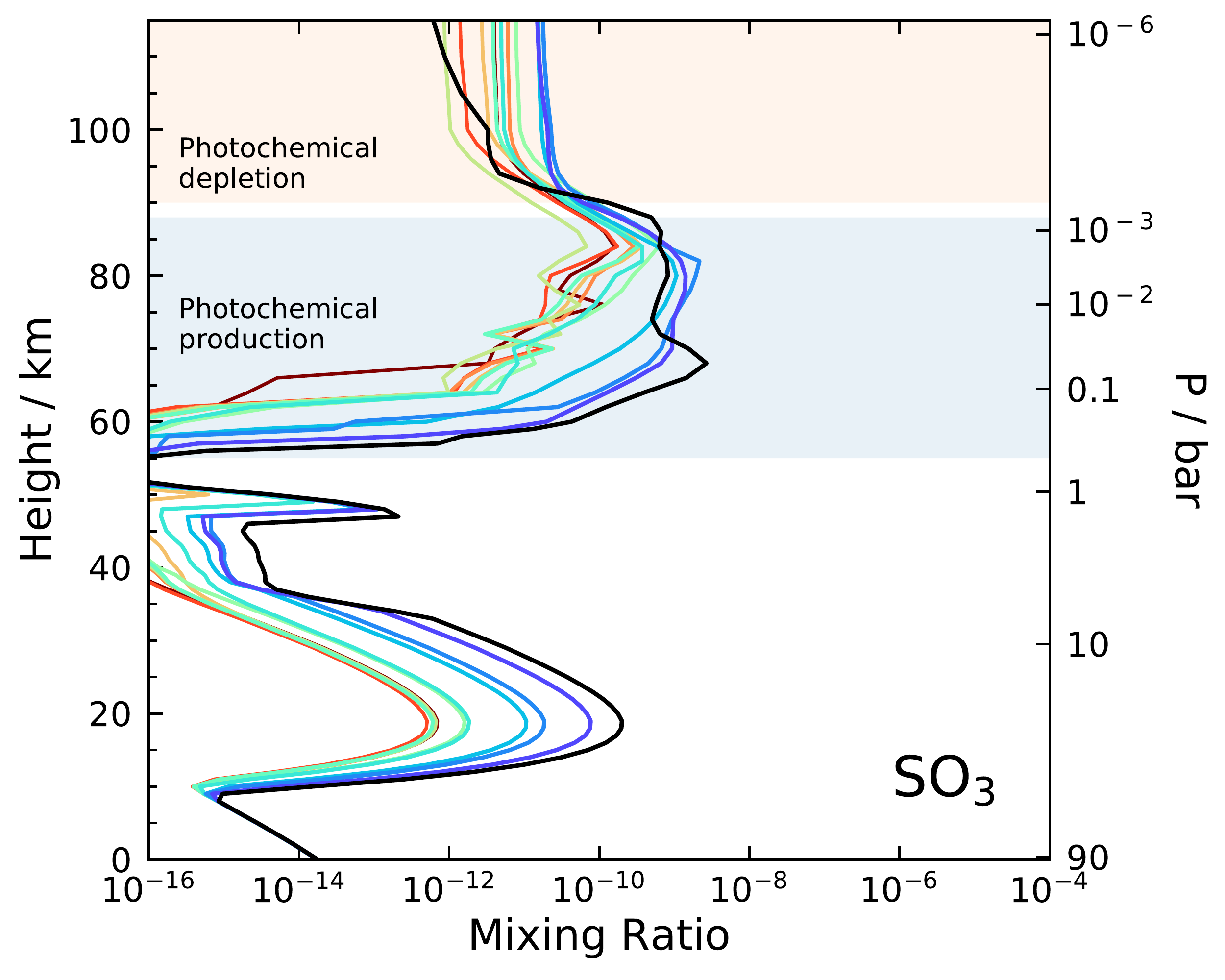}{0.45\textwidth}{}
          }
\gridline{\fig{legend.pdf}{0.8\textwidth}{}
          }
\caption{Mixing ratio of \ce{SO} (\textit{left}), and \ce{SO3} (\textit{right}) as a function of height and pressure. Droplet chemistry is included. \label{fig:SO_SO3_cloud_chem}}
\end{figure*}

On Venus, the production rate of \ce{SO} is exceeded by its photochemical destruction rate above $90\,{\rm km}$ and the \ce{SO} mixing ratio drops by three orders of magnitude (Figure \ref{fig:SO_SO3_cloud_chem}). For the K-Dwarf and M-Dwarf host stars, both the rate of reaction \ref{eq:SO2_hv} producing \ce{SO} above $60\,{\rm km}$, and the rate of \ce{SO} depletion above $90\,{\rm km}$, decrease with decreasing stellar effective temperature compared to the rates of these reactions on Venus. As a result, the total above cloud column density of \ce{SO} doesn't vary significantly as stellar effective temperature decreases, however the top of atmosphere mixing ratio becomes up to three orders of magnitude greater than on Venus.

The \ce{O} atom liberated from reaction \ref{eq:SO2_hv} can react with a further \ce{SO2} molecule to form \ce{SO3} by
\begin{align}
\ce{SO2}  + \ce{O} + \ce{M} &\rightarrow \ce{SO3} + \ce{M}.
\label{eq:SO2_O}
\end{align}
On Venus, the production rate of \ce{SO3} is exceeded by its photochemical destruction rate above $90\,{\rm km}$ and the mixing ratio drops by three orders of magnitude, shown in Figure \ref{fig:SO_SO3_cloud_chem}. For the K-Dwarf and M-Dwarf host stars, the production rate of \ce{SO3} decreases for lower stellar effective temperature (with the exception of the hotter three K-Dwarfs around $80\,{\rm km}$ altitude) and \ce{SO3} is still depleted above $90\,{\rm km}$, although to a lesser extent than on Venus. The lower \ce{SO3} production, particularly between $60$ -- $70\,{\rm km}$, impacts the production of \ce{H2SO4} molecules. \ce{H2SO4} molecules condense and form the clouds on Venus, therefore this decrease in \ce{SO3} production decreases the available cloud mass on Venus-like exoplanets around cool stars.

\begin{figure*}[ht!]
\gridline{\fig{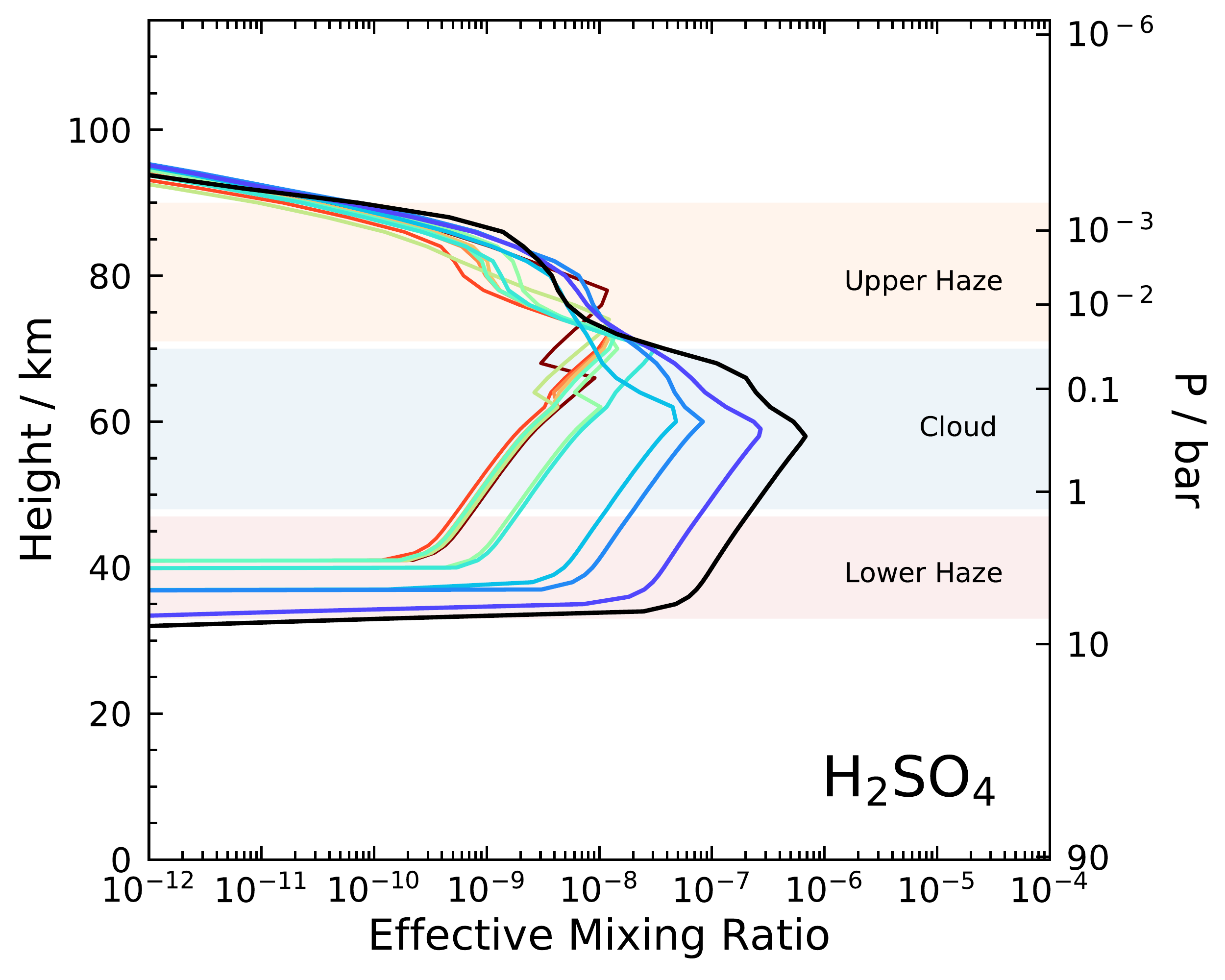}{0.8\textwidth}{}
          }
\gridline{\fig{legend.pdf}{0.8\textwidth}{}
          }
\caption{Mixing ratio as a function of height and pressure for condensed \ce{H_2SO_4}. This is an `effective' mixing ratio because the species is condensed into the liquid phase. In reality these molecules would nucleate into aerosol droplets. The profile roughly coincides with the observed cloud and haze layers on Venus \citep{Titov2018}. Droplet chemistry is included. \label{fig:H2SO4_cloud_chem}}
\end{figure*}

In the cloud layer, \ce{H2SO4 (l)} is formed by the reaction of \ce{H2O} with \ce{SO3} by
\begin{align}
\ce{SO3}  + \ce{H2O} + \ce{H2O} &\rightarrow \ce{H2SO4} + \ce{H2O}.
\label{eq:SO3_H2O}
\end{align}
On Venus the rate of this reaction peaks around $\sim 65\,{\rm km}$ altitude and the \ce{H2SO4} molecules condense into cloud droplets that settle lower in the atmosphere. The model \ce{H2SO4 (l)} profile, shown with a black line in Figure \ref{fig:H2SO4_cloud_chem}, evaporates around $35\,{\rm km}$ altitude at the lower haze boundary. This is similar to, but slightly discrepant with, the true clouds of Venus where the \ce{H2SO4 (l)} droplets begin to evaporate at $\sim50\,{\rm km}$ altitude. At lower altitudes the molecules thermally decompose back into \ce{H2O} and \ce{SO3} and this is seen in the model results with an increase in \ce{SO3} mixing ratio near $20\,{\rm km}$ in Figure \ref{fig:SO_SO3_cloud_chem}. The cloud layer is a central feature of the Venusian atmosphere, providing a sink for sulfur species and water, and leading to the observational degeneracies of Venus-like exoplanets (Section \ref{sec:intro}). 

For the K-Dwarf host stars, the \ce{H2SO4 (l)} mixing ratio decreases in order of decreasing stellar effective temperature, and for the M-Dwarf host stars, the \ce{H2SO4 (l)} mixing ratios are consistently around two orders of magnitude less than the \ce{H2SO4 (l)} mixing ratio on Venus, shown in Figure \ref{fig:H2SO4_cloud_chem}. This follows directly from the lesser production of \ce{SO3}. We discuss the potential impact of this result further in Section \ref{sec:discussion}.

\subsection{Oxygen cycles} 
\label{sec:oxygen_cycles}

In addition to production from \ce{SO2} photolysis, atomic oxygen is also produced from the photochemistry of \ce{CO2} and this is the primary pathway to producing \ce{O} atoms above $\sim70\,{\rm km}$ in the atmosphere:
\begin{align}
\ce{CO2}  + h\nu &\rightarrow \ce{CO} + \ce{O}.
\label{eq:CO2_hv}
\end{align}
The oxygen atoms that react with an \ce{SO2} molecule form \ce{SO3} and ultimately end up in the \ce{H2SO4 (l)} cloud layer. However, this is the path for a minority of the \ce{O} atoms produced via photochemistry. The fate of the majority of the atomic oxygen depends on the relative balance of catalytic cycles involving photochemical products from the other parent species in the atmosphere: \ce{HO_x}, \ce{ClO_x}, \ce{ClCO_x}, \ce{SO_x}, \ce{HSO_x} and \ce{NO_x} \citep{YungDeMore1982}. These species are families of oxides where the subscript \ce{x} denotes the number of oxygen atoms in the molecule. They are short lived in the atmosphere and are capable of catalysing the formation of \ce{O2} from \ce{O} atoms, the reformation of \ce{CO2} from \ce{O} and \ce{CO}, or the destruction of \ce{O2}.

Photochemical \ce{O2} has not been observed in the upper atmosphere of Venus, leading to an inferred upper limit on its abundance of $<2.8\,{\rm ppm}$ \citep{Marcq2018}. Although much less abundant in the model predictions, \ce{O3} has been detected \citep{Montmessin2011, Marcq2019}. \ce{O3} is a photochemical byproduct of \ce{O2} in an atmosphere and is treated as a tracer of \ce{O2}. Given the upper limit on \ce{O2} concentration in the Venus atmosphere, the detections of $550\,{\rm ppb}$ \ce{O3} at $\sim 105\,{\rm km}$ altitude \citep{Montmessin2011} and $10.8\,{\rm ppb}$ \ce{O3} at $\sim 65\,{\rm km}$ altitude \citep{Marcq2019}, suggest that \ce{O3} is being formed via unknown chemical pathways. The detection of \ce{O3} and non-detection of \ce{O2} on Venus occur at different latitudes and so an alternative interpretation might be that there is potentially a source of spatially localised \ce{O2} on Venus, however, since there is no geographical overlap between the \ce{O2} and \ce{O3} observations, no meaningful comparison can be drawn. Whatever the true state of oxygen is on Venus, no $1{\rm D}$ photochemical models to date are able to destroy \ce{O2} efficiently enough to be consistent with its inferred upper limit, including the model in this study. The inability of this, or any, photochemical model to accurately predict the \ce{O2} abundance on Venus generates significant uncertainty in all model results that rely upon oxygen, both molecular and atomic, and even more uncertainty when extrapolating to exoplanets. The predicted \ce{O2} profile for Venus using this model is nonetheless in closer agreement with the observed upper limit compared to past photochemical models \citep[see e.g.,][]{Zhang2012,BiersonZhang2020}.

\begin{figure*}[ht!]
\gridline{\fig{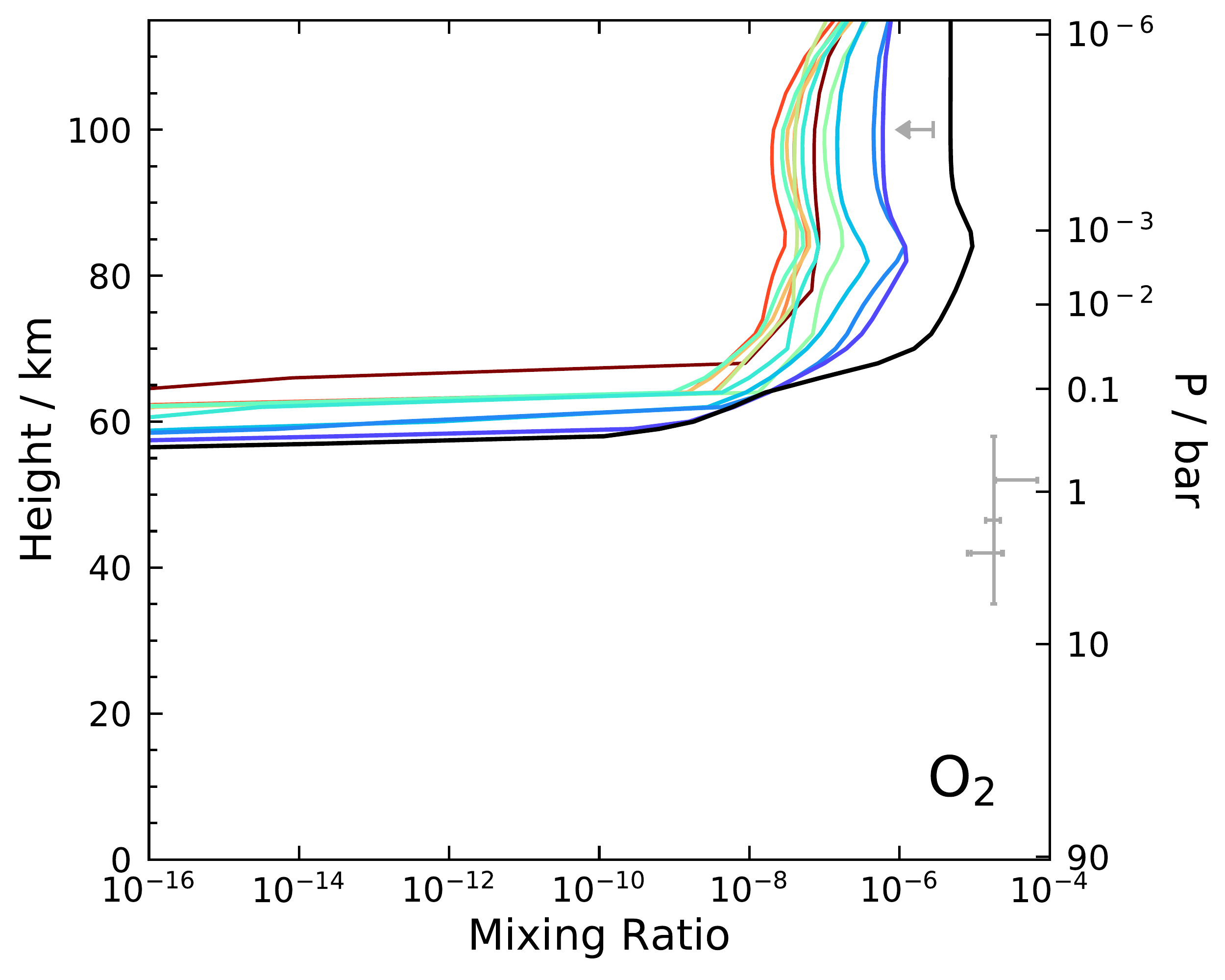}{0.45\textwidth}{}
          \fig{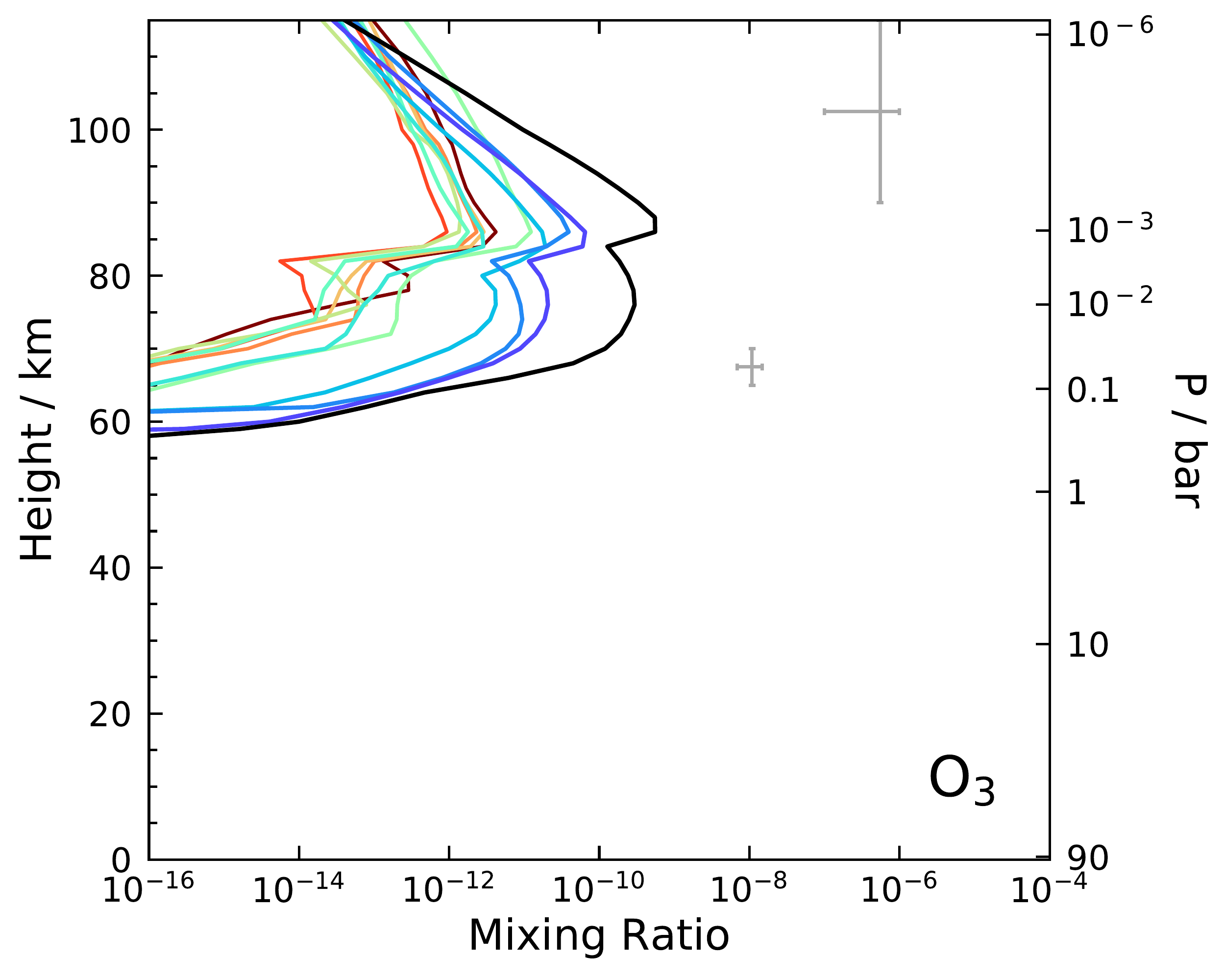}{0.45\textwidth}{}
          }
\gridline{\fig{legend.pdf}{0.8\textwidth}{}
          }
\caption{Mixing ratio of \ce{O2} (\textit{left}), and \ce{O3} (\textit{right}) as a function of height and pressure. Droplet chemistry is included. \label{fig:O2_O3_cloud_chem}}
\end{figure*}

The abundance of photochemical oxygen and ozone produced in the model is shown in Figure \ref{fig:O2_O3_cloud_chem}. The mixing ratio of \ce{O2} on Venus is a factor of $\sim2$ above the inferred upper limit, while the mixing ratio of \ce{O3} on Venus lies well below the observed values. For the K-Dwarf and M-Dwarf host stars the mixing ratios of \ce{O2} and \ce{O3} decrease with decreasing stellar effective temperature. \ce{O2} reaches a peak mixing ratio near $\sim85\,{\rm km}$ for Venus and this peak gradually disappears with decreasing stellar effective temperature. The ozone mixing ratio exhibits two peaks, one sharp peak near $\sim85\,{\rm km}$ which remains present for all the stellar spectra, and a broader peak near $\sim75\,{\rm km}$ which does not appear with the lower effective temperature M-Dwarf fluxes.

The main reaction cycle forming \ce{O2} at its peak near $\sim85\,{\rm km}$ is catalysed by \ce{HO_x} radicals, sourced from \ce{H2O} photolysis:
\begin{align}
\ce{H}  + \ce{O2} + \ce{M} &\rightarrow \ce{HO2} + \ce{M},\\
\ce{HO2}  + \ce{O} &\rightarrow \ce{HO} + \ce{O2},\\
\ce{HO}  + \ce{O} &\rightarrow \ce{H} + \ce{O2},\\
Net:  \ce{O}  + \ce{O} &\rightarrow \ce{O2}.
\label{eq:H_O2}
\end{align}

\ce{HO_x} radicals are also responsible for reforming \ce{CO2} from \ce{CO} and \ce{O} and therefore the relative balance of these catalytic pathways determines the precise shape of the \ce{O2} profile. Nonetheless, uncatalysed \ce{O2} formation from \ce{O} atoms becomes the dominant formation pathway between $\sim90$ -- $105\,{\rm km}$:
\begin{align}
\ce{O}  + \ce{O} + \ce{M} &\rightarrow \ce{O2} + \ce{M}
\label{eq:O_O}
\end{align}

Thus the overabundance of \ce{O2} remains a problem that requires an efficient mechanism for \ce{O2} destruction regardless of the balance of \ce{HO_x} catalysed pathways. The cycle catalysed by \ce{SO_x} is the most dominant destruction pathway between $\sim70$ -- $95\,{\rm km}$:
\begin{align}
\ce{S}  + \ce{O2} &\rightarrow \ce{SO} + \ce{O},\\
\ce{SO}  + h\nu &\rightarrow \ce{S} + \ce{O}, \\
Net:  \ce{O2} &\rightarrow \ce{O}  + \ce{O}.
\label{eq:S_O2}
\end{align}

In addition to this, \citet{YungDeMore1982} suggest that \ce{ClCO3} catalysts might play an important role in \ce{O2} destruction and bring the \ce{O2} abundance in photochemical models of Venus below the inferred upper limit, via
\begin{align}
\ce{Cl}  + \ce{CO} + \ce{M} &\rightarrow \ce{ClCO} + \ce{M},\\
\ce{ClCO}  + \ce{O2} + \ce{M} &\rightarrow \ce{ClCO3} + \ce{M},\\
\ce{ClCO3}  + \ce{X} &\rightarrow \ce{Cl} + \ce{CO2} + \ce{XO}, \\
Net:  \ce{CO} + \ce{O2} + \ce{X} &\rightarrow \ce{CO2} + \ce{XO},
\label{eq:ClCO_O2}
\end{align}
\noindent where \ce{X = H, Cl, O, SO, or SO2}. This is capable of destroying \ce{O2} when \ce{X = SO_x} and when \ce{X = H} if the \ce{HO} produced goes on to react with \ce{CO} rather than \ce{O}. This route has the advantage of being catalysed by \ce{ClCO_x} species without requiring a photon in the reaction cycle, and thus only requires photochemical liberation of \ce{Cl} from \ce{HCl} to be initiated. However we find that the formation rate of \ce{ClCO3} is two to three orders of magnitude below any other destruction pathway for \ce{O2}. \ce{ClCO3} thus has a negligible impact on \ce{O2} abundance in this study. \ce{Cl} catalysed pathways are also of lesser importance than \ce{H} catalysed pathways for \ce{O2} formation, are of comparable importance for \ce{CO2} reformation, and the majority of \ce{H} that participates in \ce{HO_x} catalysed cycles originates from \ce{H2O} rather than \ce{HCl}.

\begin{figure*}[ht!]
\gridline{\fig{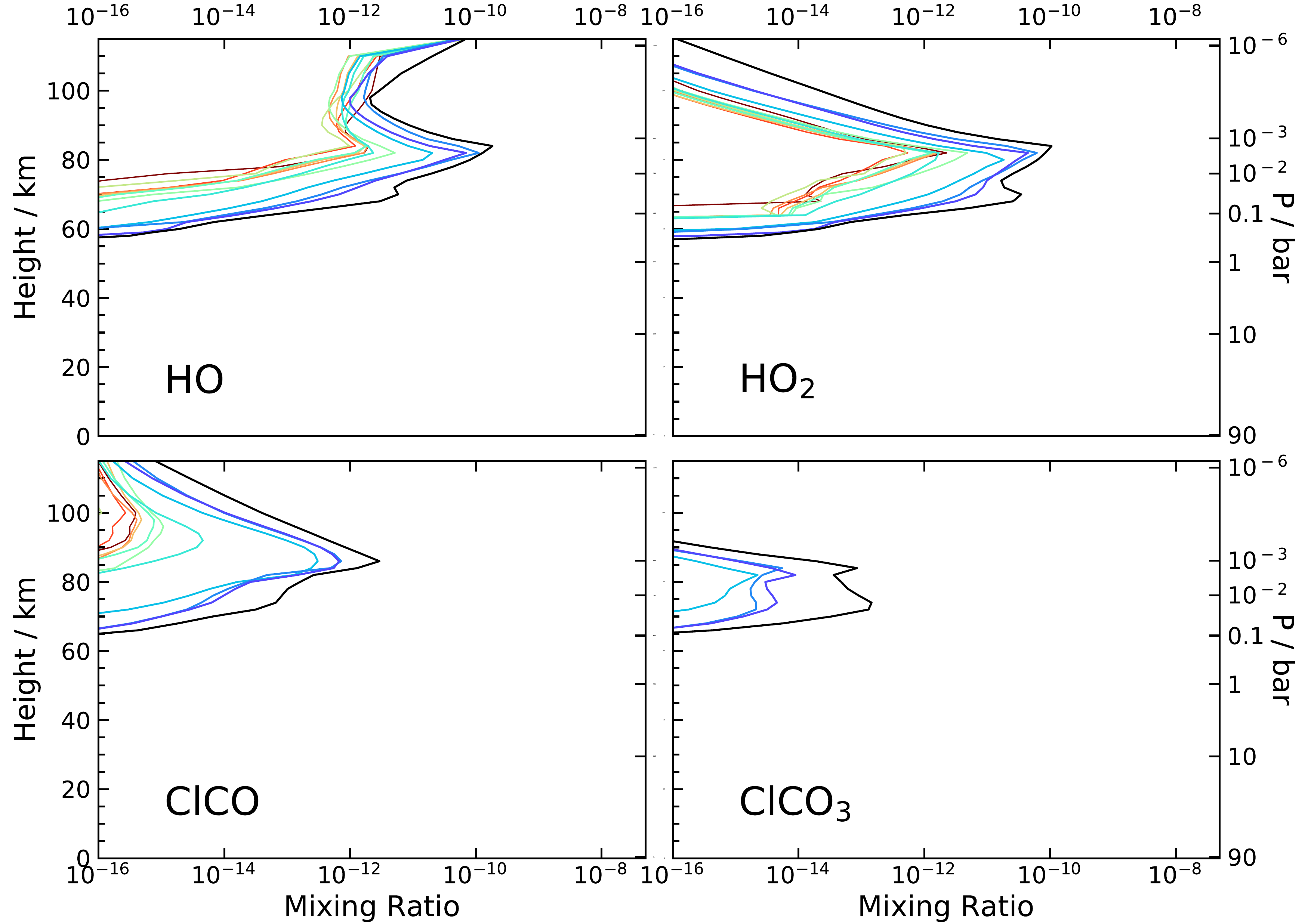}{\textwidth}{}
          }
\gridline{\fig{legend.pdf}{0.8\textwidth}{}
          }
\caption{Mixing ratios of \ce{HO}, \ce{HO2}, \ce{ClCO}, and \ce{ClCO3} as a function of height and pressure. Droplet chemistry is included. \label{fig:radicals_cloud_chem}}
\end{figure*}

Figure \ref{fig:radicals_cloud_chem} shows the mixing ratios of the radical species \ce{HO}, \ce{HO2}, \ce{ClCO} and \ce{ClCO3} for the different stellar spectra. The abundances of all the catalysts decrease with decreasing stellar effective temperature, however the \ce{Cl} species diminish far more so than the \ce{H} species. For the lowest effective temperature K-Dwarf and all of the M-Dwarf host stars, the action of chlorine catalysts in both \ce{O2} formation and \ce{CO2} formation becomes negligible compared to that of the hydroxyl catalysts, meanwhile \ce{ClCO3} is not produced at all and \ce{S} atoms remain the most important catalysts for destroying \ce{O2}.

\section{Discussion} 
\label{sec:discussion}

\subsection{Observational discriminants of Venus-like worlds}
\label{sec:observational_discriminants}

\ce{SO2}, \ce{OCS} and \ce{H2S} are spectroscopically active species. They are heavily depleted above the clouds of Venus and would make poor observational indicators of Venus analogue exoplanets around Sun-like stars. However, Section \ref{sec:phot_parent} shows that they are not photochemically depleted by the flux from the coolest K-Dwarf and the seven M-Dwarf host stars in this study. Provided these species endure in sufficient abundance below the clouds of a Venus analogue around a cool K-Dwarf or quiescent M-Dwarf host star, they will provide a set of observable indicators accessible with near future telescopes. \ce{SO2} has IR bands at $7.3$, $8.7$, and $19.3\,{\rm \mu m}$, \ce{OCS} has IR bands at $4.8$, $11.6$, and $19.1\,{\rm \mu m}$, and \ce{H2S} has IR bands at $3.8$, $4.2$, and $8.5\,{\rm \mu m}$ \citep{Fegley2014}. These spectral features are within the observing range of JWST ($0.6$ -- $28\,{\rm \mu m}$). However, the question of how high their abundances must be for a sufficient detection threshold is more difficult to predict.

Detection of trace sulfur gases in transmission spectroscopy has been explored in past work with synthetic spectra of Venus and Venus-like exoplanets. These synthetic spectra have usually been compared with synthetic spectra of Earth-like exoplanets in order to find observational discriminants between the two paradigms. \citet{KalteneggerSasselov2010} have produced synthetic spectra of Earth-like atmospheres with and without artificially high levels of \ce{SO2} and \ce{H2S}. They identified strong features in the \ce{SO2} IR bands $7.3$, $8.7$, and $19.3\,{\rm \mu m}$ for the models with $1\,{\rm ppm}$ \ce{SO2}, that were distinguishable from the models without. \citet{Hu2013} found that diagnostic \ce{SO2} absorption features at $7.3$ and $19.3\,{\rm \mu m}$ were detectable for mixing ratios above $0.1\,{\rm ppm}$. Absorption in the IR bands of \ce{H2S} were weaker and the only feature they generated was a pseudo-continuum absorption at wavelengths beyond $30\,{\rm \mu m}$, coinciding with the rotational bands of \ce{H2O}. This feature was only detectable in models that were highly desiccated and with high sulfur emission rates compared to the Earth.

In contrast, \citet{Barstow2016} generated synthetic spectra of Earth and Venus analogue exoplanets and found that an Earth-like atmosphere provided the best fit to both spectra when a reduced-cloud a priori model was used to fit the cloudy-Venus spectrum. The reason for this was a lack of observable species unique to Venus. Fitting the modern Earth-analogue spectrum was unambiguous due to a strong ozone absorption feature at $9\,{\rm \mu m}$ whereas the Venus analogue produced a near-flat spectrum with no distinguishing features except for a \ce{CO2} absorption feature at $4.3\,{\rm \mu m}$ that was also present in the Earth spectrum. The \ce{SO2} feature at $8.7\,{\rm \mu m}$ was seen but only on the order of the noise in their synthetic data. However, a complication to exploiting ozone as a discriminant comes from \citet{Ehrenreich2012} whose synthetic spectrum of Venus transiting the Sun showed that the Hartley band ozone absorption emerged from Mie scattering when the vertical localisation of the Venusian ozone layer was considered \citep{Montmessin2011}. A further complication to using ozone as a discriminant comes from the fact that the Earth's atmosphere has not always contained significant levels of ozone over its geological history \citep{Catling2014}.

The takeaway from these past studies is that conclusions drawn from synthetic spectra are highly dependent on the assumed abundances of species, the assumed vertical distribution of species, and modelling approaches used. The results presented here favour the observability of \ce{SO2}, \ce{OCS} and \ce{H2S} on Venus-analogues around cool K-Dwarf and M-Dwarf hosts, however it remains difficult to place a definitive threshold on the minimum required mixing ratio for such a detection, even before any assumption about aerosol scattering or JWST noise floors are considered. It is possibly the case that the above-cloud abundances of \ce{SO2} and \ce{OCS} are just high enough in our M-Dwarf model results to be detectable, and \ce{H2S} will not be, while none of the three species would be detectable for the reference Venus model or the hotter K-Dwarfs. We do not attempt to produce synthetic spectra in this study and so this conclusion remains to be tested further.

There is also observational evidence that the \ce{OCS} mixing ratio decreases from $30$ -- $40\,{\rm km}$ altitude which is not reproduced by our model \citep{Pollack1993}. This decrease in \ce{OCS} is observed to be anticorrelated with \ce{CO} and points to a reaction converting \ce{OCS} into \ce{CO} below the clouds. We do find this conversion occurring in our model but it is much less pronounced than the observational trend, and \ce{OCS} then reforms above $40\,{\rm km}$ altitude back to its original abundance, before being photochemically depleted above $50\,{\rm km}$. It is unclear whether the above-cloud mixing ratio of \ce{OCS} would be substantially different if our model were able to reproduce the observations for below-cloud \ce{OCS}, and future work is required to correctly model these observational data.

\subsection{Pressure sensitivity}
\label{sec:pressure_sensitivity}

\begin{figure*}[ht!]
\gridline{\fig{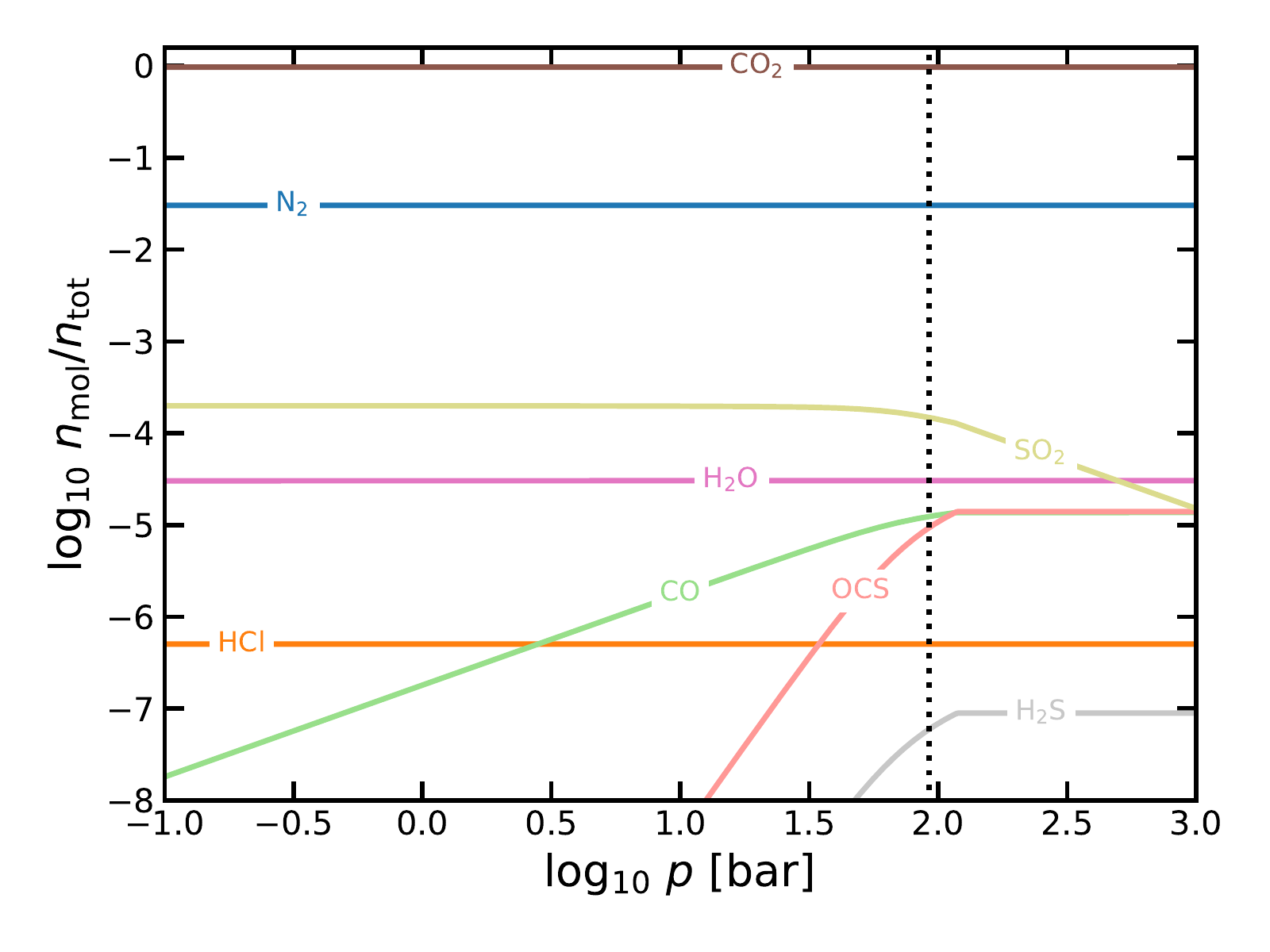}{0.8\textwidth}{}
         }
\caption{{\sc GGchem} model for the bottom of the Venus atmosphere assuming thermochemical equilibrium between the surface and atmosphere, as a function of surface pressure, at a fixed temperature of $735\,{\rm K}$. \label{fig:pressure}}
\end{figure*}

In general for Venus-like exoplanets the surface pressure could also vary, and this would alter the initial chemical abundances input at the base of the model atmospheres. \citet{Rimmer2021} show that the initial chemical abundances from table \ref{tab:init} are consistent with the interpretation that the surface and atmosphere of Venus are in thermochemical equilibrium. The analysis was done by calculating equilibrium abundances of gas and condensate phases at the surface temperature and pressure of Venus based off the Vega $2$ measurements of oxide ratios at the surface, using the program {\sc GGchem} \citep{Woitke2018}. Here we briefly examine how the initial species abundances may be expected to vary with surface pressure for a set of Venus-like exoplanets whose surfaces and atmospheres are assumed to be in thermochemical equilibrium. We follow the same procedure using {\sc GGchem} as that outlined in \citet{Rimmer2021} to fit for the surface-atmosphere composition of Venus at $735\,{\rm K}$ and $92\,{\rm bar}$ pressure. We then explore how the abundances of the initial species from table \ref{tab:init} vary with pressure over the range $0.1 - 1000\,{\rm bar}$, shown in figure \ref{fig:pressure}.

At surface pressures lower than Venus, the abundances of \ce{OCS} and \ce{H2S} drop off most rapidly with decreasing pressure, while the abundance of \ce{CO} drops off more gradually. At surface pressures higher than Venus, only the \ce{SO2} abundance varies, decreasing with increasing pressure. \ce{CO2}, \ce{N2}, \ce{H2O}, and \ce{HCl} abundances remain constant across the pressure range as these species can be fitted directly with their prescribed gas phase abundances in {\sc GGchem}, without any possible mineral phases to condense into at $735\,{\rm K}$ \citep{Rimmer2021}.

This result is relevant to \ce{OCS}, \ce{SO2} and \ce{H2S} as potential observational discriminants of Venus-like worlds, with both \ce{OCS} and \ce{H2S} abundances diminishing within surface pressures $\sim1$ order of magnitude lower than that of Venus. Despite this, across the whole pressure range at least one of \ce{SO2} and \ce{OCS} remain at comparable abundances to the reference Venus atmosphere and could still be potential observables in the upper atmosphere. Nonetheless, surface pressure and temperature may not be the whole story. It could be the case that the atmosphere is set by volcanic degassing which will likewise have its own pressure, temperature, and compositional profile, and will substantially complicate the story for potential Venus-like exoplanets.

\subsection{Varying below-cloud \ce{SO2}}
\label{sec:vary_so2}

The enhanced survival of sulfur gases on Venus-like worlds around cooler host stars has so far only been shown for true Venus-analogues, with the same abundance of \ce{SO2}, \ce{OCS} and \ce{H2S} at the surface as on Venus. The timescale for removing \ce{SO2} entirely from the atmosphere of Venus, via chemical weathering of Calcium-bearing minerals at the surface, is on the order of Myr \citep{Fegley2014}. The observation of abundant \ce{SO2} therefore indicates that there is a flux of \ce{SO2} from the surface maintaining it as the third most abundant atmospheric constituent, or that the current `\ce{SO2}-rich' state of Venus is only transient. Furthermore, in section \ref{sec:pressure_sensitivity} we show that for varying surface pressure, the \ce{SO2} abundance at the base of the Venus atmosphere remains constant for lower pressures but decreases with increasing pressure. The relatively high abundance of atmospheric \ce{SO2} is therefore not necessarily a universal property of Venus-like exoplanets. Here, we assess how robust our conclusion of enhanced survival of sulfur gases with reducing stellar effective temperature is in more diverse exoplanet contexts, by investigating how the above-cloud survival of \ce{SO2}, \ce{SO}, \ce{OCS} and \ce{H2S}, is affected by the surface \ce{SO2} abundance. Reducing the below-cloud \ce{SO2} in our model represents a broader range of Venus-like exoplanets, with lower \ce{SO2} content due to either different volcanic activity, surface geochemistry, or bulk composition.

\begin{figure*}[ht!]
\gridline{\fig{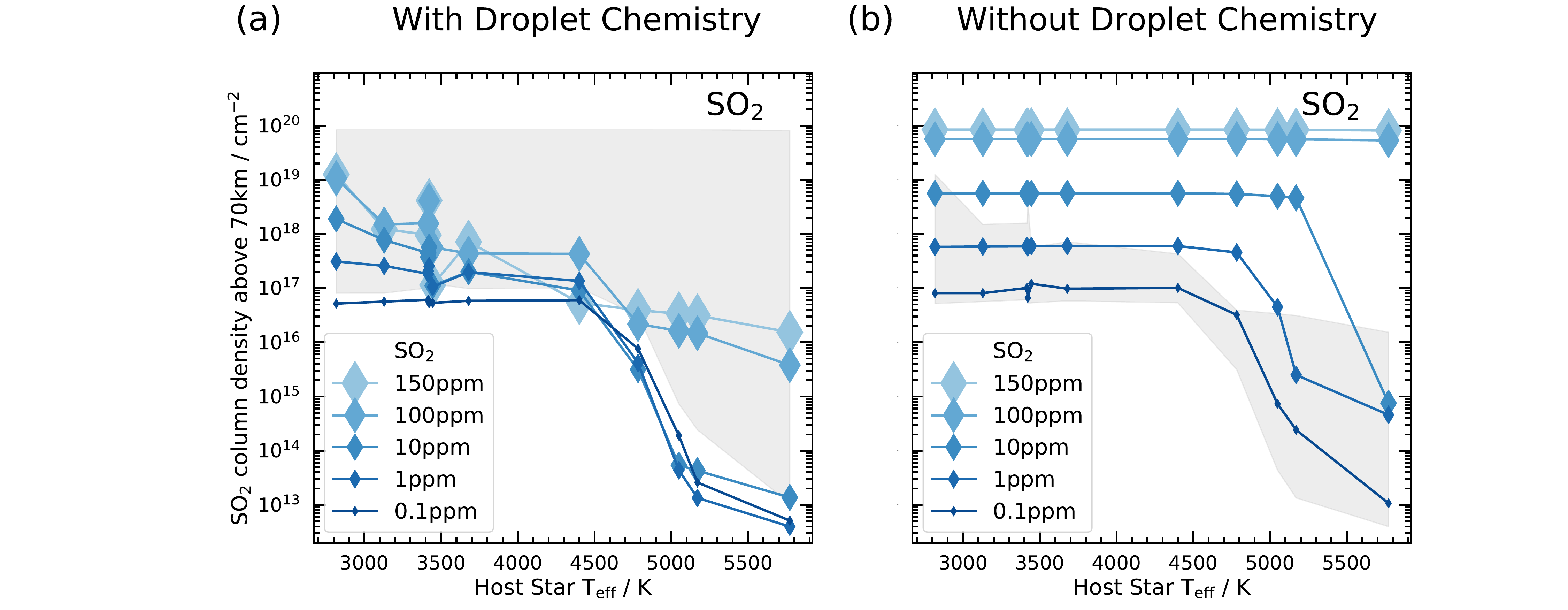}{\textwidth}{}
          }
\gridline{\fig{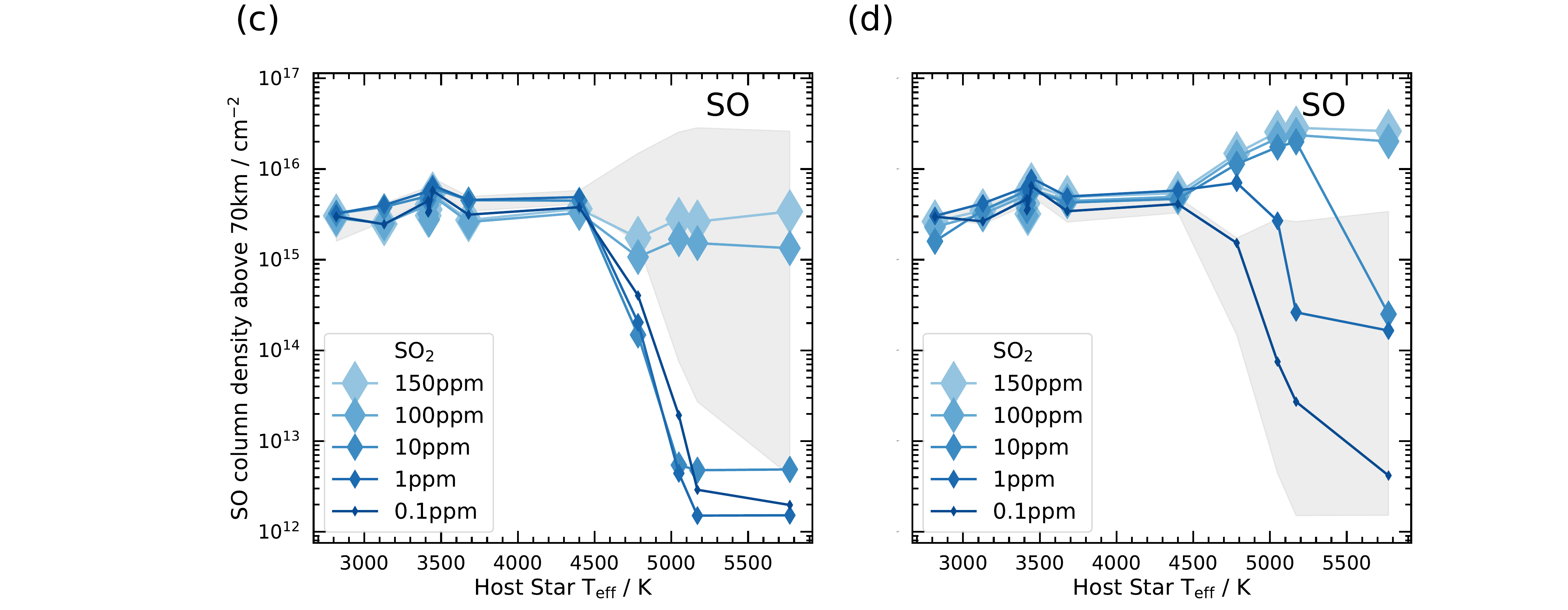}{\textwidth}{}
          }
\caption{Integrated column densities above the cloud top (assumed fixed at $70\,{\rm km}$) for \ce{SO_2} and \ce{SO}, as the below-cloud \ce{SO2} mixing ratio is varied from $150\,{\rm ppm}$ down to $0.1\,{\rm ppm}$. In the left hand column results with droplet chemistry included are plotted in colour and the shaded region behind shows the bounds of the results without droplet chemistry. In the right hand column the results without droplet chemistry are shown plotted in colour and the shaded region behind shows the bounds of the results with droplet chemistry.  \label{fig:vary_SO2}}
\end{figure*}

Figure \ref{fig:vary_SO2} shows how the column density of \ce{SO2}, \ce{SO}, \ce{OCS} and \ce{H2S} molecules varies as a function of stellar effective temperature of the host star, for a range of surface \ce{SO2} values. We estimate observable column density by integrating the number density from the top of the atmosphere down to $70\,{\rm km}$ altitude, where the Venusian cloud top lies. The left hand column of Figure \ref{fig:vary_SO2} shows the results with droplet chemistry included and the shaded region behind shows the bounds of the results without droplet chemistry, and vice versa for the right hand column.

\begin{figure*}[ht!]
\gridline{\fig{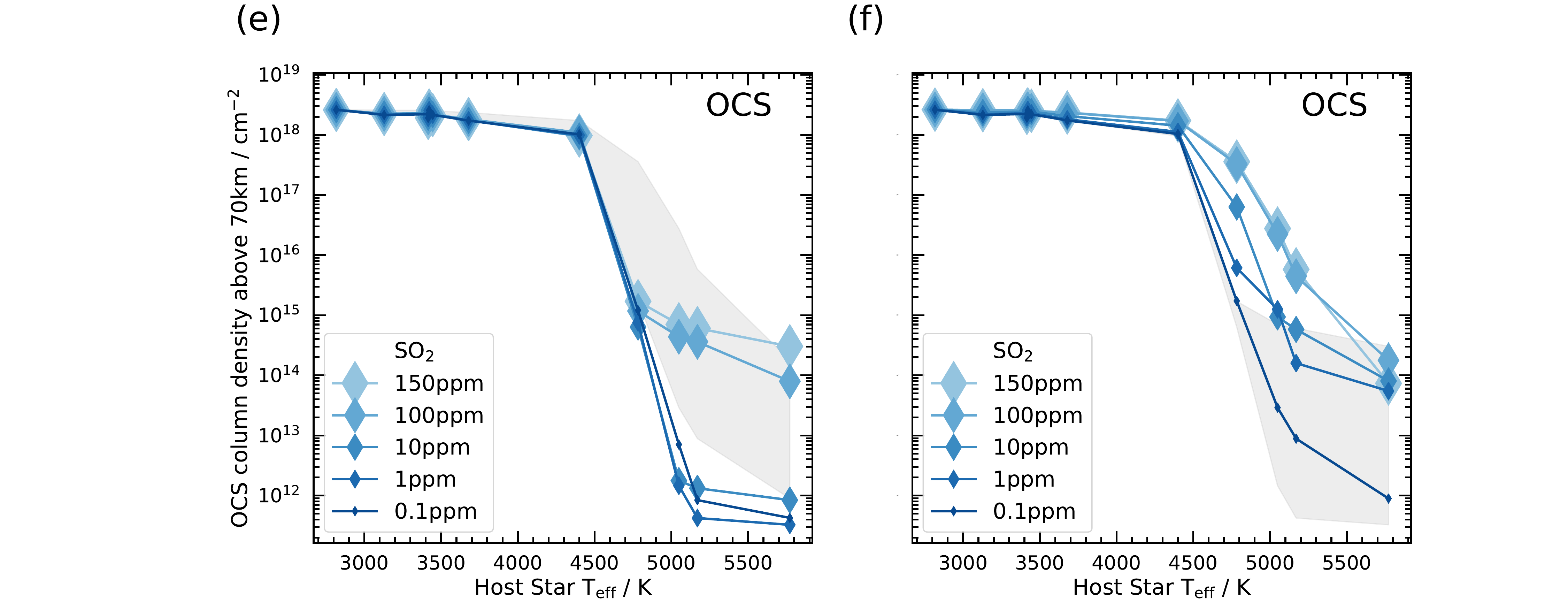}{\textwidth}{}
          }
\gridline{\fig{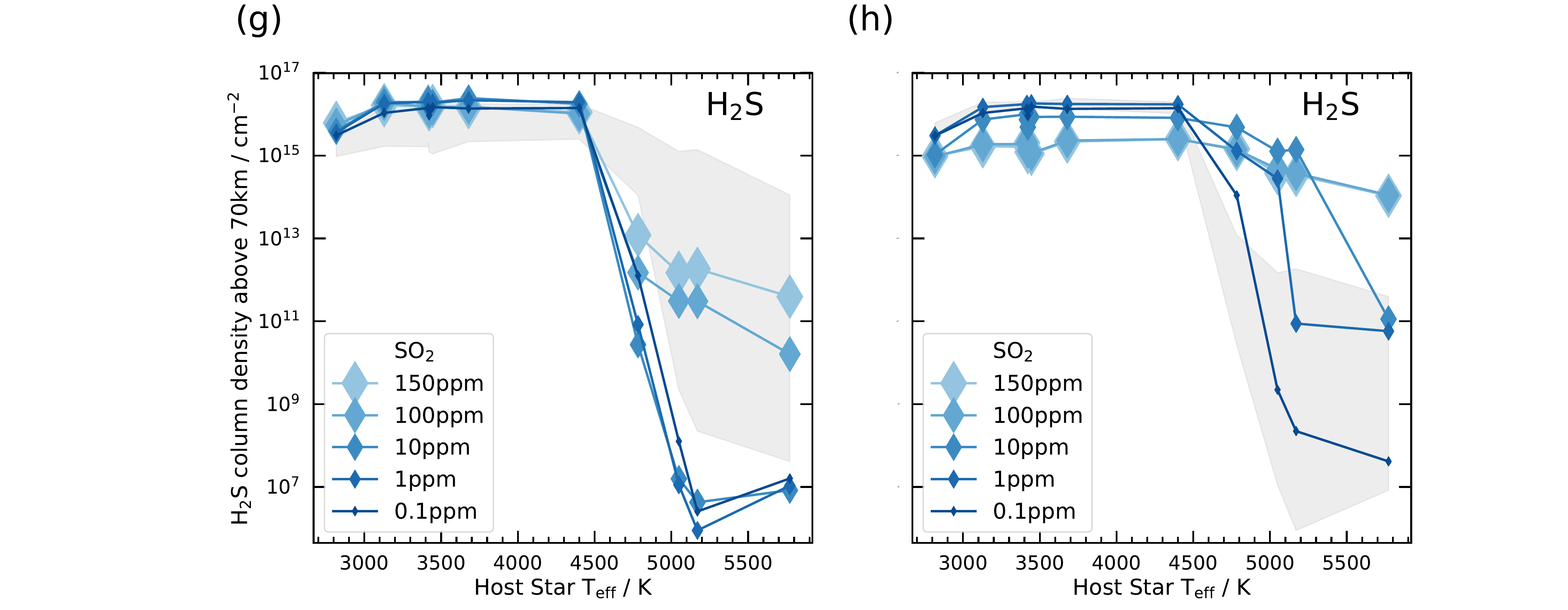}{\textwidth}{}
          }
\caption{Same as figure \ref{fig:vary_SO2} for \ce{OCS} and \ce{H2S} column densities above the cloud top.  \label{fig:vary_SO2_2}}
\end{figure*}

Figure \ref{fig:vary_SO2} shows that the enhanced survival of trace sulfur gases above the clouds of Venus-like exoplanets is robust to order of magnitude differences in the below-cloud abundance of \ce{SO2}, when host star T$_{\rm{eff}}\,<\,4500\,{\rm K}$, with droplet chemistry occurring in the clouds. Without droplet chemistry, the enhanced survival of \ce{OCS} is still robust, however a divergence in \ce{SO2}, \ce{SO} and \ce{H2S} column density with increasing host star T$_{\rm{eff}}$ only emerges for surface \ce{SO2} around an order of magnitude less than on Venus. Of course it should also be noted that if droplet chemistry is not operating on a Venus-analogue exoplanet, and there is no other mechanism of \ce{SO2} depletion apart from photochemistry at the top of the atmosphere, then the lack of in-cloud \ce{SO2} depletion would result in observable levels of \ce{SO2} anyway. 

These results are promising for utilising spectral features of trace sulfur gases as observational discriminants of Venus-like exoplanets around cool K-Dwarfs and quiescent M-Dwarfs, particularly for \ce{OCS} which maintains the same mixing ratio above the clouds as it has below the clouds for the coolest K-Dwarf and all the M-Dwarf spectra in our investigation, independently of droplet chemistry. This result however comes with the caveat that there is potentially significant \ce{OCS} to \ce{CO} conversion below the clouds which is not reproduced by our model to the same extent as some observations suggest, as discussed in Section \ref{sec:observational_discriminants}.

\subsection{The Cloud Layer}
\label{sec:cloud_layer}

The results of Section \ref{sec:sulfur_species} showed that the cloud layer of \ce{H2SO4 (l)} molecules decreases for decreasing stellar effective temperature, shown in Figure \ref{fig:H2SO4_cloud_chem}. This is also true for the models without droplet chemistry as shown in Figure \ref{fig:H2SO4_no_cloud_chem} in the Appendix. This trend follows directly from the different production rates of \ce{SO3}, and when droplet chemistry is included, the resulting \ce{H2SO4 (l)} profile aligns well with the observed boundaries of the cloud and haze layers on Venus.

As a model of the cloud layer, Figure \ref{fig:H2SO4_cloud_chem} is crude and cannot be related to aerosol nucleation or droplet size distributions, but the result nonetheless suggests that there will be a smaller reservoir of cloud-forming material for Venus-like exoplanets hosted by cooler stars. This result is independent of droplet chemistry, but is more pronounced when droplet chemistry is included because the cloud top occurs at lower altitude and the \ce{H2SO4 (l)} profile has a lesser vertical extent. With the M-Dwarf stellar spectra, all of the \ce{H2SO4 (l)} mixing ratios through the main cloud deck are over two orders of magnitude less than on Venus, and, if one assumes a fixed droplet size distribution, then this implies a lesser number density of aerosol droplets populating the cloud layer, which would decrease the opacity and enable observations to probe to lower altitudes. Combining this with the result that trace sulfur bearing gases are not photochemically depleted by the M-Dwarf fluxes then means that their potential observability on a Venus-like exoplanet around a quiescent M-Dwarf host star is yet more favourable. However aerosol nucleation and optical properties are complex. The clouds of Venus are observed to have three different aerosol modes in the main cloud deck \citep{Titov2018} and it could still be the case that the abundance of \ce{H2SO4 (l)} has a non-linear effect on the different particle size distributions of the aerosol modes. Modelling of aerosol nucleation would be required in order to make concrete predictions about the optical effects of the changing \ce{H2SO4 (l)} profile.

Clouds also have a profound effect on a planet's climate. On Venus the albedo of the cloud layer reflects over $75\,{\rm \%}$ of the incoming Solar flux while also having a powerful greenhouse effect trapping thermal radiation at wavelengths $\lambda\,>\,2.5\,{\rm\mu m}$ \citep{Titov2018}. The changing number density of aerosols in the cloud layer would likely change the radiative balance and ultimate surface temperature of the planet, potentially in highly non-linear ways, and investigating these effects would require detailed cloud and climate modelling that is beyond the scope of this study. Allowing the temperature structure of the lower atmosphere to vary would also lead to feedback mechanisms between the climate and the cloud layer as the altitudes of \ce{H2SO4 (l)} evaporation and thermal decomposition shift higher or lower. This would be an interesting future avenue for investigation, with the possibility of coupling thermochemistry of mineral buffers at the surface to determine the chemical mixing ratios at the base of the atmosphere as the surface temperature changes.

\subsection{Abiotic Oxygen}
\label{sec:abiotic_oxygen}

In Section \ref{sec:oxygen_cycles} we found that the pathway catalysed by hydroxyl, \ce{OH}, was the most active catalytic cycle forming \ce{O2} on Venus and Venus-like exoplanets. Hydroxyl is also responsible for catalysing reformation of \ce{CO} and \ce{O} into \ce{CO2}, in combination with \ce{ClCO} catalysts. The balance of the hydroxyl-catalysed reactions converting \ce{O} into \ce{O2} versus \ce{CO} and \ce{O} into \ce{CO2} determines the fate of the atomic oxygen liberated by photochemistry. Hydroxyl preferentially reforms \ce{O} into \ce{CO2} between $\sim60$ -- $75\,{\rm km}$ and above $\sim105\,{\rm km}$, while preferentially forming \ce{O2} instead between $\sim75$ -- $105\,{\rm km}$, and this is responsible for the peak in \ce{O2} mixing ratio around $85\,{\rm km}$ in Figure \ref{fig:O2_O3_cloud_chem}. 

It has previously been found that the hydroxyl-catalysed cycles reforming \ce{CO2} are operating in the Martian atmosphere, with the hydroxyl radicals originating from photolysis of \ce{H2O} \citep{McElroyDonahue1972, Nair1994}. As the atmosphere of Venus is much drier than the atmosphere of Mars it has often been argued that chlorine catalysts from \ce{HCl} are more important for \ce{CO2} stability and \ce{O2} formation/destruction than \ce{HO_x} catalysts from \ce{H2O} \citep{YungDeMore1999, MillsAllen2007}. The results from this investigation do not support this, with the chlorine-catalysed cycles being of lesser importance in \ce{O2} formation/destruction than the \ce{HO_x} pathways, and of comparable importance to \ce{HO_x} for \ce{CO2} stability only between $\sim85$ -- $105\,{\rm km}$. 

We also found that the \ce{O2} destruction pathway via \ce{ClCO3} is negligible in the Venus atmosphere despite suggestions that it may be the solution to the \ce{O2} overabundance problem found in previous photochemical models of Venus \citep{YungDeMore1982}. This result is consistent with the findings from \citet{Kras2012}, although our model results for \ce{O2} are nonetheless closer to the observed limit. Further to this, for the Venus-analogue exoplanets, as the stellar effective temperature decreased, we found that the rate of chlorine-catalysed pathways all decreased disproportionately to the hydroxyl-catalysed pathways. These results suggest that the presence of \ce{HCl} at its nominal abundance on Venus-like exoplanets around cool stars has little effect on the atmospheric \ce{O2} and \ce{CO2}. 

As stellar effective temperature decreased, the \ce{O2} and \ce{O3} mixing ratios decreased, particularly so at lower altitudes where the atmosphere is thicker, suggesting that steady-state photochemical oxygen on Venus-like exoplanets is less likely to be observed. While the mixing ratio of ozone in our model does not meet the observed values from both \citet{Montmessin2011} at $\sim105\,{\rm km}$ and \citet{Marcq2019} at $\sim65\,{\rm km}$, the shape of the vertical distribution in Figure \ref{fig:O2_O3_cloud_chem} can be seen to separate into two discrete peaks, matching the observation of two individual layers, as also seen in the modelling from \citet{YungDeMore1999} \citep{Montmessin2011}. For the Venus-analogue exoplanets, as the stellar effective temperature of the host star decreases, the ozone mixing ratio decreases by up to five orders of magnitude around $\sim60$ -- $80\,{\rm km}$, and decreases less so for increasing altitudes. The peak near $\sim85\,{\rm km}$ remains present as the lower altitude layer diminishes, and the mixing ratio at the very top of the atmosphere doesn't change. Whilst the higher altitude layer is better positioned to allow spectroscopic investigation of the atmosphere in transit, the lower altitude layer has higher number density due to the increased pressure.

This would be a promising result for reducing the potential false-positive biosignature of oxygen on Venus-like exoplanets, however three quite serious limitations apply to these inferences. The first is that we do not match observations of \ce{O2} and \ce{O3} on Venus, nor does any photochemical model of the full-atmosphere. While it is possible that there is a spatially localised source of \ce{O2} on Venus responsible for the \ce{O3} detections from \citet{Montmessin2011} and \citet{Marcq2019}, there is still no explanation as yet for the non-detection of \ce{O2} near equatorial regions suggesting efficient destruction via an unknown mechanism. We therefore cannot account for how this mechanism might vary for Venus-like exoplanets. As future models come closer to solving the mystery of \ce{O2} and \ce{O3} on Venus it will be important to re-examine the atmospheric processes at work in an exoplanetary context.

The second limitation is that the desiccation of a planetary atmosphere upon entering a runaway greenhouse phase of evolution, followed by \ce{H2O} photolysis and \ce{H}-loss to space, has been shown to result in significant \ce{O2} buildup in the atmosphere, and this planetary evolution is expected to be common for exoplanets around M-Dwarf host stars \citep{LugerBarnes2015}. This is an independent mechanism of abiotic oxygen production to that examined in our analysis and could potentially be a systematic false-positive on exoplanets. Venus nonetheless provides a counter example to this hypothesis as it also experienced desiccation and \ce{H}-loss earlier in its evolution upon entering a runaway-greenhouse phase, evidenced by the high deuterium to hydrogen ratio measured by \citet{Donahue1982} and \citet{McElroy1982}. If the mechanism of past \ce{O2} removal from the atmosphere involved something peculiar about Venus' cloud, surface, or atmospheric chemistry, then it may be that the lack of \ce{O2} on Venus cannot be generalised to exo-Venus. This makes understanding the oxygen chemistry on Venus essential for any study of false-positive \ce{O2}/\ce{O3} biosignatures.

The final limitation is that M-Dwarfs can often be highly active, as discussed in Section \ref{sec:intro}, with flaring events and coronal mass ejections occurring more frequently than for Sun-like stars \citep{Gunther2020}, resulting in a variable UV input into a planets atmosphere which will significantly alter the photochemical cycles that occur \citep{SchaeferFegley2011}. This transient photochemistry is not currently captured in our model and will depend in detail on the energies and occurrence rates of the flaring activities. We therefore ignore such effects for the purposes of this investigation and restrict the scope of our results to Venus-like exoplanets around relatively quiescent K- and M-Dwarf host stars. It remains to be determined in the future how our results will be affected for more active host stars.

\section{Conclusion} 
\label{sec:conclusion}

Venus-like exoplanets may be cosmically ubiquitous but will be difficult to distinguish from potentially habitable Earth-like planets. This observational degeneracy is caused by the global cloud layer, and the only atmospheric region that will be accessible in transmission spectroscopy, the region above the cloud layer, is dominated by photochemical processes. In this study we have investigated how these photochemical processes change for Venus-like exoplanets around K- and M-Dwarf host stars, using a validated atmospheric model of the full Venus-atmosphere from $0$ -- $115\,{\rm km}$, and stellar spectra from the MUSCLES Treasury Survey.

Our results show that \ce{SO2}, \ce{OCS}, \ce{H2S}, and \ce{SO} are not photochemically depleted above the cloud layer for the coolest K-Dwarf and all seven M-Dwarf host star fluxes, whereas these species are heavily photochemically depleted on Venus by the Solar flux. This result is robust to decreasing the surface \ce{SO2} abundance by many orders of magnitude, but does have some dependence on the inclusion of hypothesised `droplet-chemistry' in the model, with the exception of \ce{OCS}, the enhanced survival of which is independent of both droplet chemistry and surface \ce{SO2} abundance. We then examine the formation of \ce{H2SO4 (l)} molecules that form the cloud layer and find that the \ce{H2SO4 (l)} abundance steadily decreases with decreasing stellar effective temperature for the four K-Dwarfs, and is consistently two order of magnitude less than the \ce{H2SO4 (l)} abundance on Venus for all seven M-Dwarfs. 

Finally, we investigate the catalytic formation/destruction of photochemical oxygen and ozone. We find that \ce{O2} builds up in the upper atmosphere primarily due to both uncatalysed formation from \ce{O} atoms, and catalysed formation via \ce{HO_x} radicals, while \ce{ClCO_x} radicals have a lesser effect on the atmospheric \ce{O2}. As stellar effective temperature of the host star was decreased, the abundances of the chlorine radicals decreased disproportionately to the hydrogen radicals until all catalytic chlorine cycles were negligible. Overall, the atmospheric \ce{O2} and \ce{O3} mixing ratios decrease with decreasing stellar effective temperature of the host stars.

Our results all point toward trace sulfur-bearing gases as being prime observational indicators of Venus-like exoplanets hosted by cool K-Dwarf or quiescent M-Dwarf host stars, with enhanced above-cloud mixing ratios and potentially lesser cloud formation compared to Venus in the Solar System. Future work that could extend this kind of analysis would be the inclusion of climate-modelling, detailed cloud-modelling, and investigation of the observed spectra that would result from these Venus-like exoplanets in transit around K- and M-Dwarf host stars.

\acknowledgments

We thank the reviewer for their detailed and considered feedback which helped us to improve the clarity of our results and the underlying science. S. J. thanks the Science and Technology Facilities Council (STFC) for a PhD studentship (grant reference ST/V50659X/1). P. B. R. thanks the Simons Foundation for funding (SCOL awards 599634).

\software{\textsc{Argo} \citep{RimmerHelling2016}, \textsc{GGchem} \citep{Woitke2018}}

\appendix

\section{Without droplet chemistry}
\label{sec:no_cloud_chem}

Here we replicate all plots with droplet chemistry not included.

\begin{figure*}[ht!]
\gridline{\fig{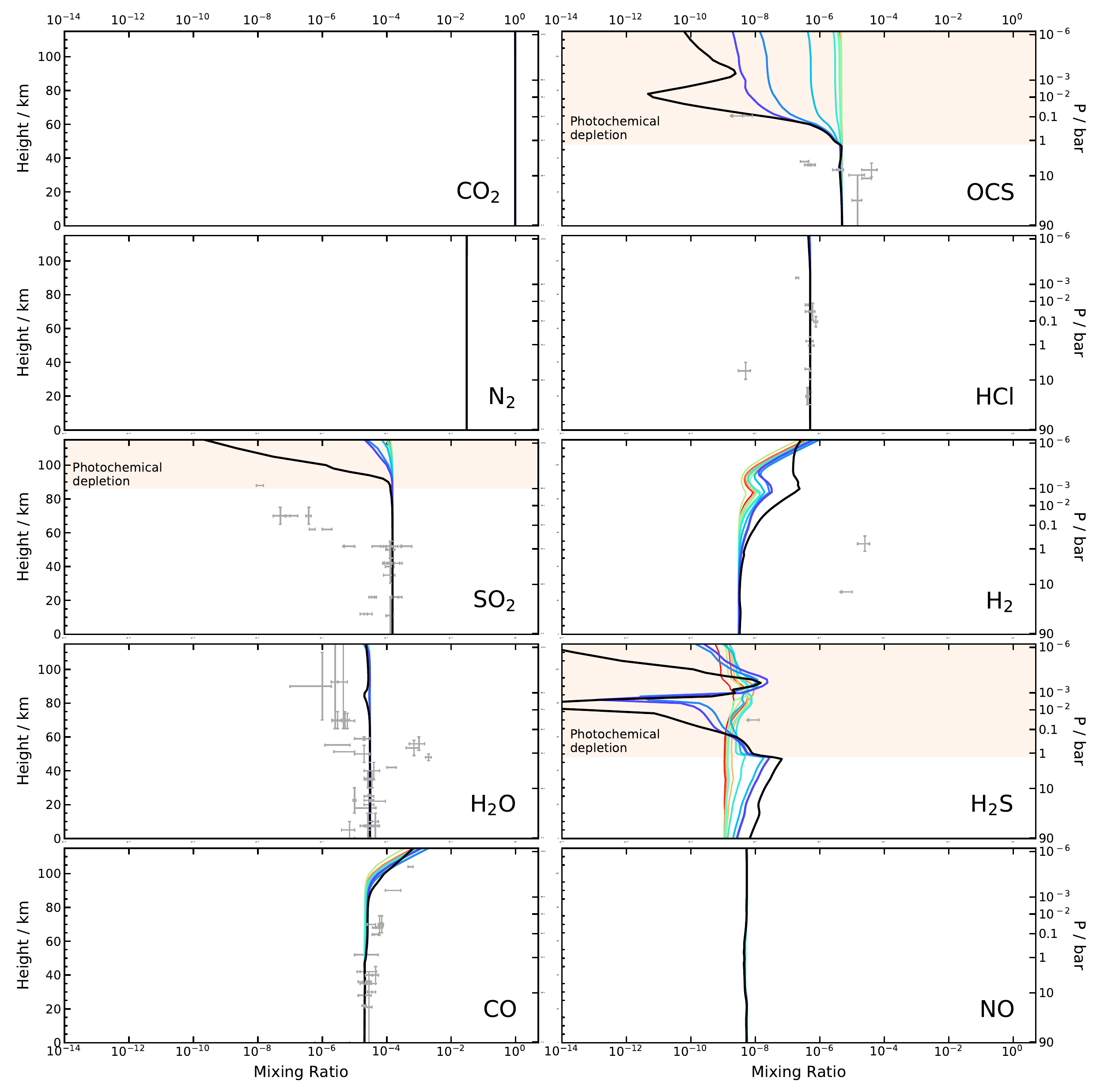}{\textwidth}{}}
\gridline{\fig{legend.pdf}{0.8\textwidth}{}}
\caption{Photochemical parent species in the atmosphere, in order of decreasing surface mixing ratio. Droplet chemistry is not included.  \label{fig:phot_parents_no_cloud_chem}}
\end{figure*}

\begin{figure*}[ht!]
\gridline{\fig{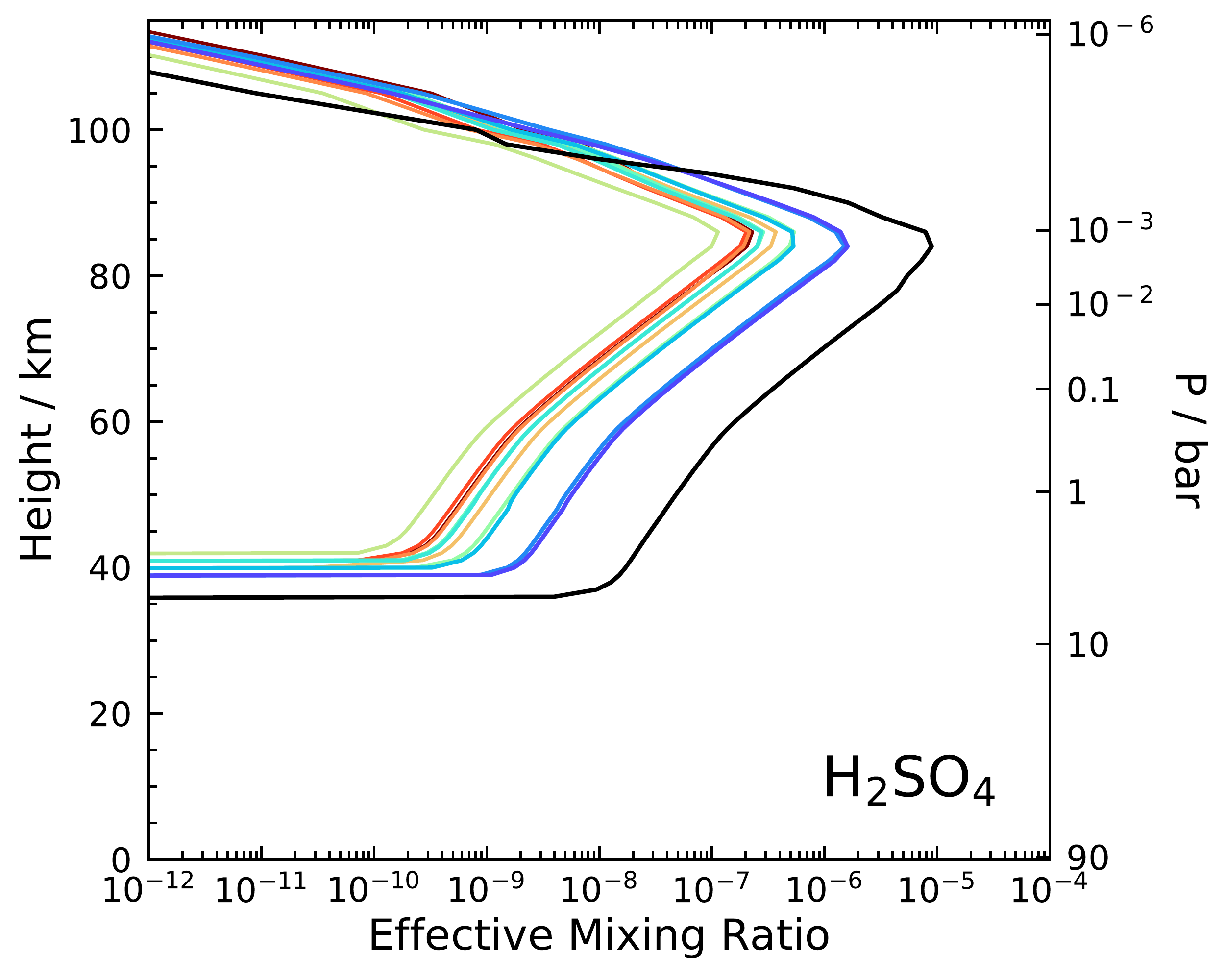}{0.8\textwidth}{}
          }
\gridline{\fig{legend.pdf}{0.8\textwidth}{}
          }
\caption{Effective mixing ratio as a function of height and pressure for condensed \ce{H_2SO_4}, with no droplet chemistry included. This is an `effective' mixing ratio because the species is condensed into the liquid phase and in reality these molecules would nucleate into aerosol droplets. The trend of decreasing \ce{H2SO4 (l)} abundance with decreasing effective temperature remains apparent. The cloud top coincides with $\sim90\,{\rm km}$, in disagreement with the observed value of $\sim70\,{\rm km}$ \citep{Titov2018}. \label{fig:H2SO4_no_cloud_chem}}
\end{figure*}

\begin{figure*}[ht!]
\gridline{\fig{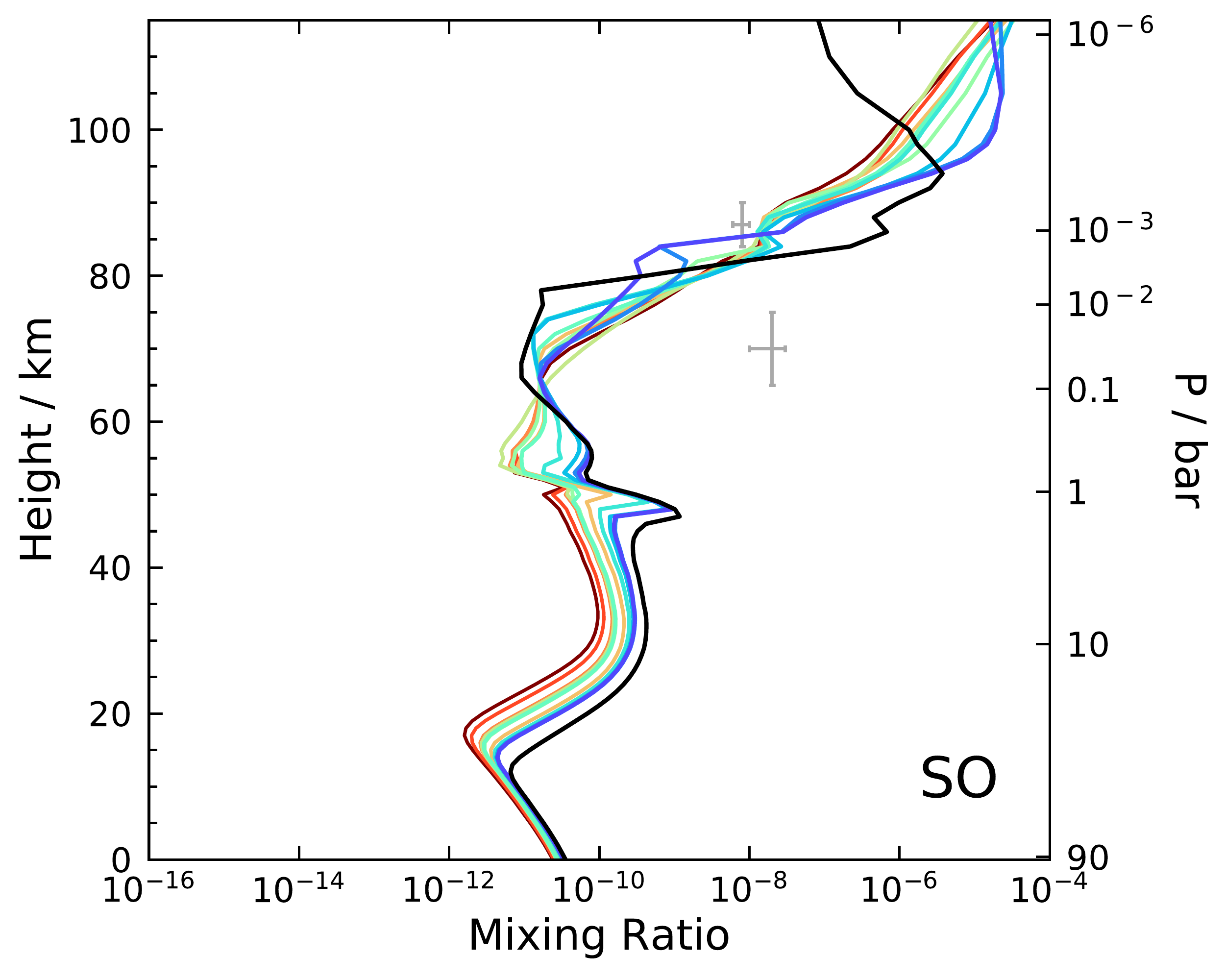}{0.45\textwidth}{}
          \fig{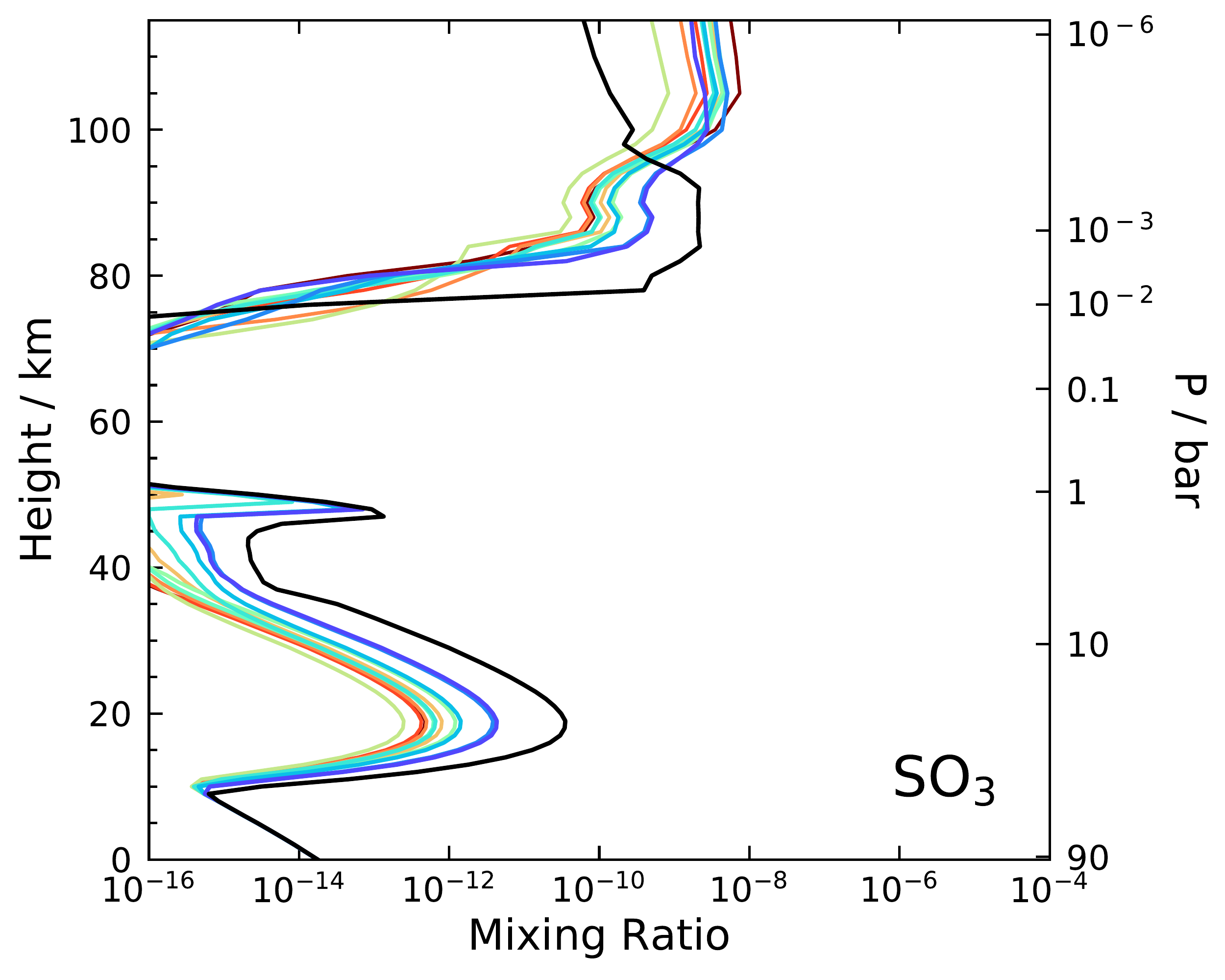}{0.45\textwidth}{}
          }
\gridline{\fig{legend.pdf}{0.8\textwidth}{}
          }
\caption{Mixing ratio of \ce{SO} (\textit{left}), and \ce{SO3} (\textit{right}) as a function of height and pressure. Droplet chemistry is not included. \label{fig:SO_SO3_no_cloud_chem}}
\end{figure*}

\begin{figure*}[ht!]
\gridline{\fig{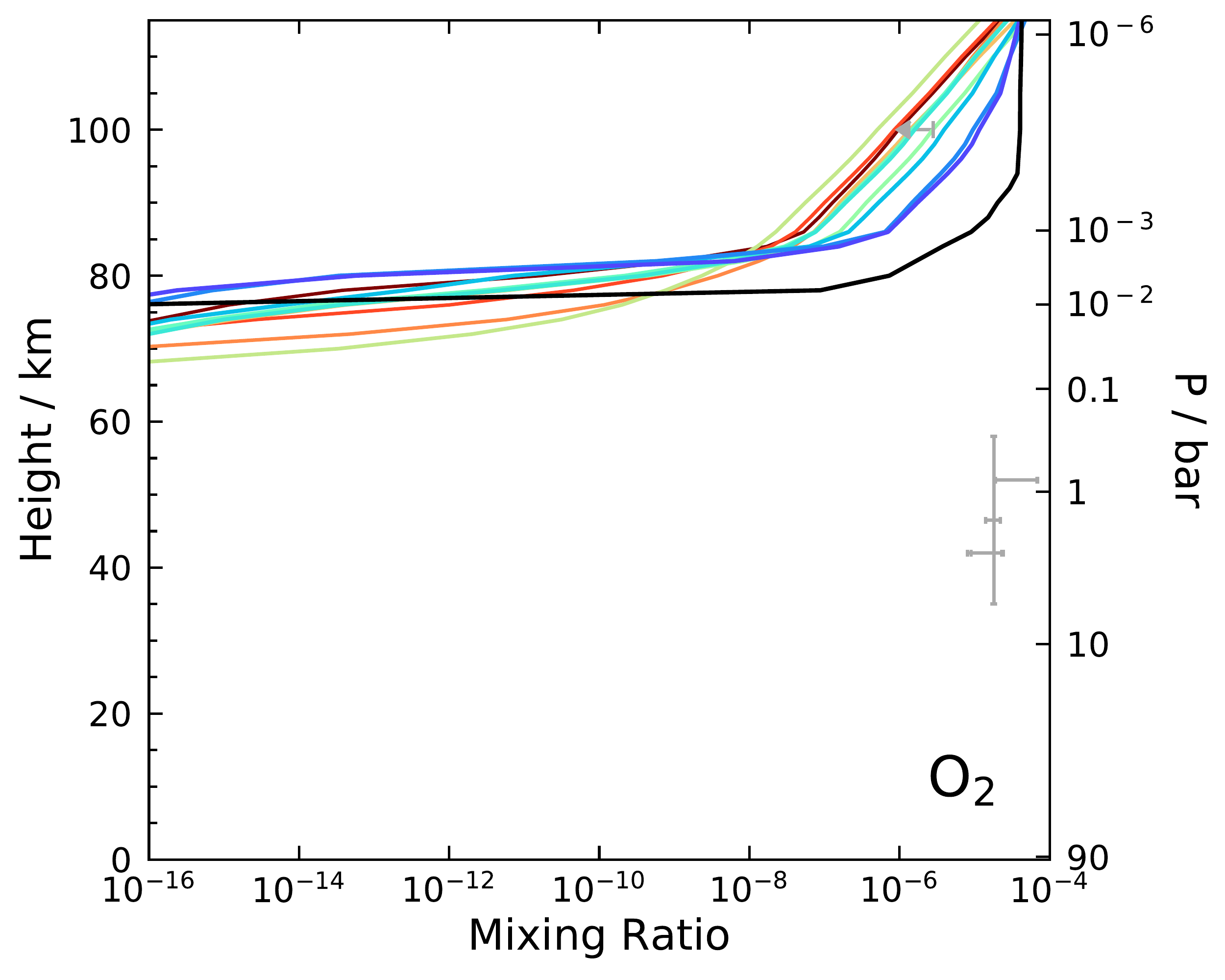}{0.45\textwidth}{}
          \fig{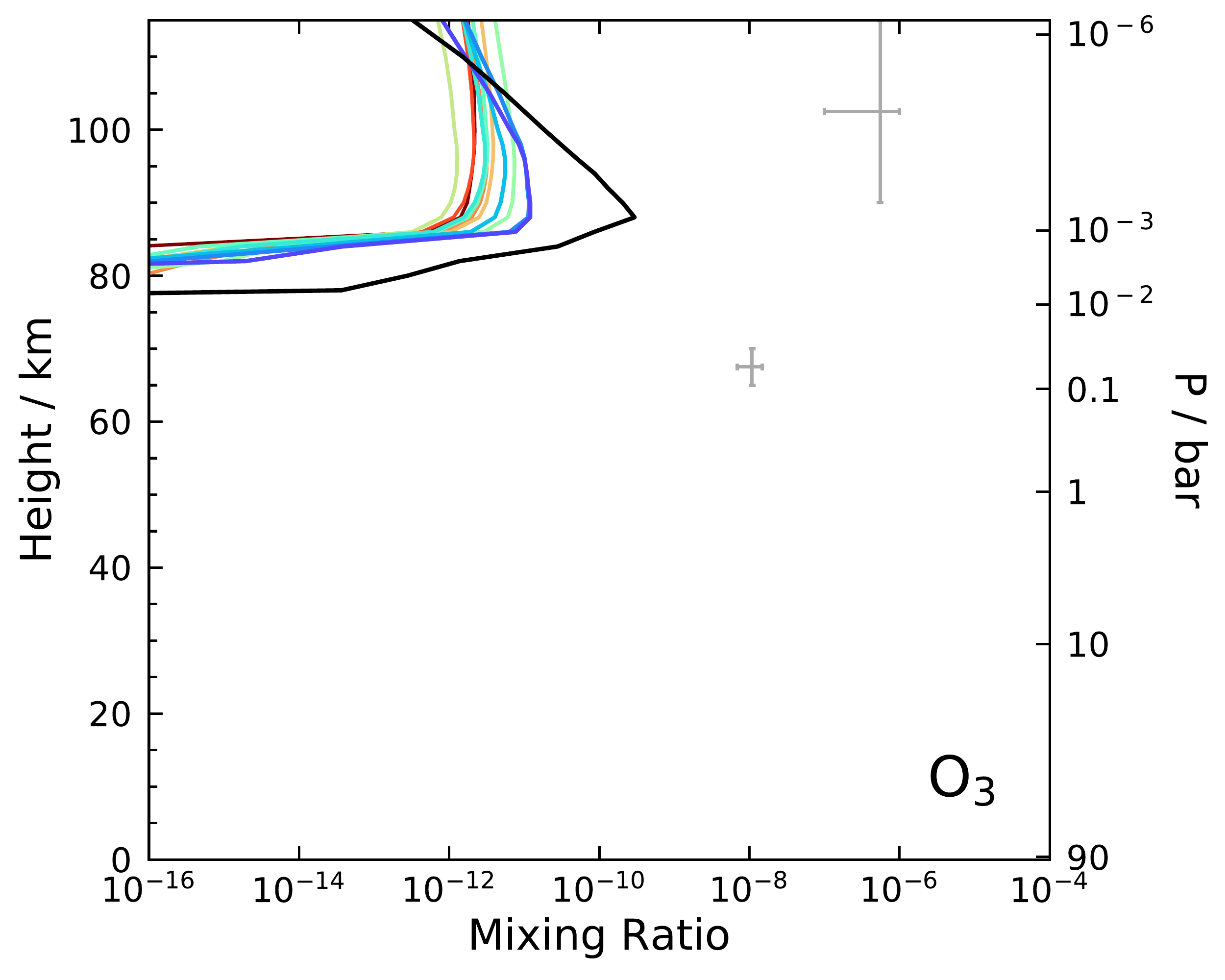}{0.45\textwidth}{}
          }
\gridline{\fig{legend.pdf}{0.8\textwidth}{}
          }
\caption{Mixing ratio of \ce{O2} (\textit{left}), and \ce{O3} (\textit{right}) as a function of height and pressure. Droplet chemistry is not included. \label{fig:O2_O3_no_cloud_chem}}
\end{figure*}

\begin{figure*}[ht!]
\gridline{\fig{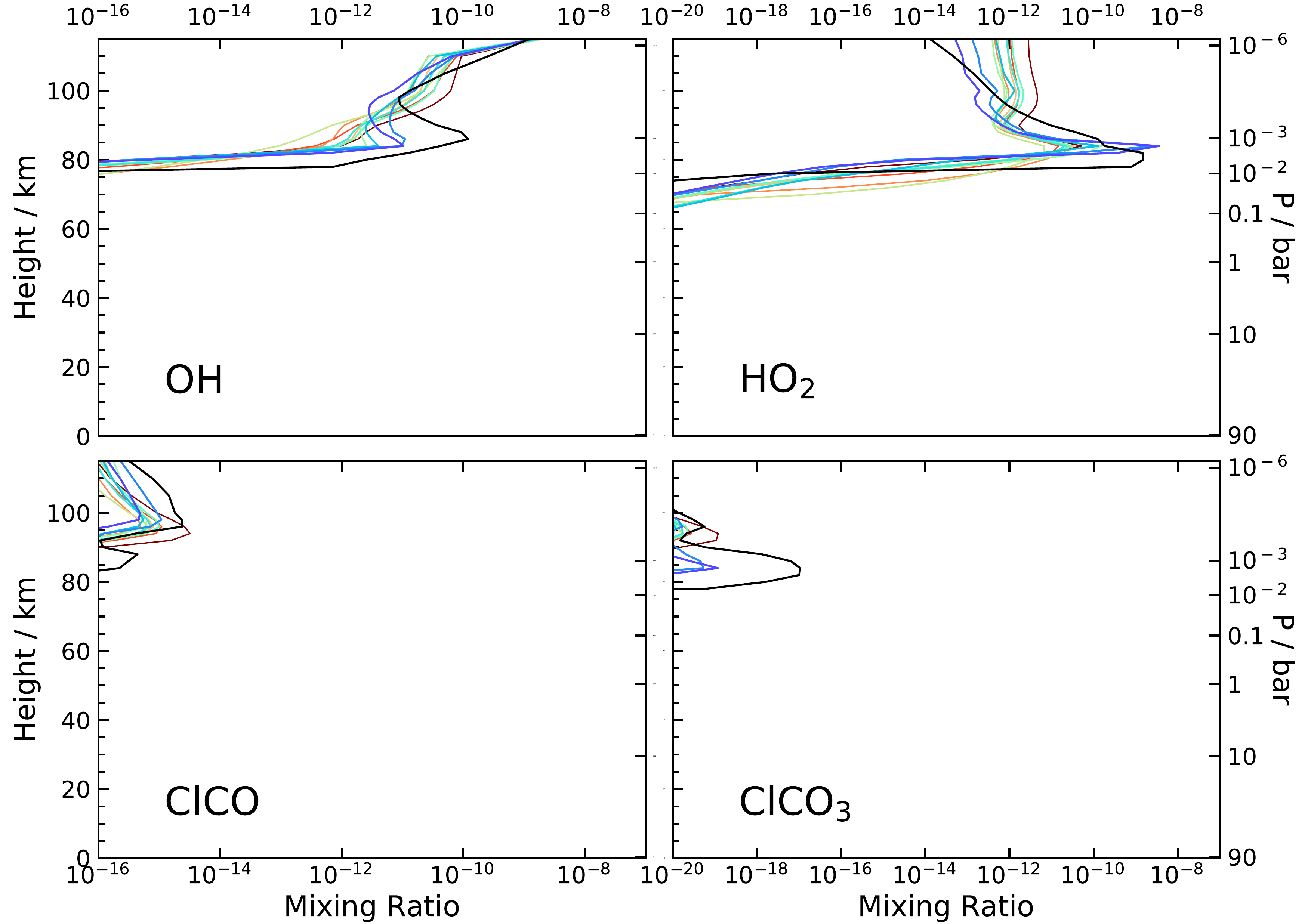}{\textwidth}{}
          }
\gridline{\fig{legend.pdf}{0.8\textwidth}{}
          }
\caption{Mixing ratios of \ce{HO}, \ce{HO2}, \ce{ClCO}, and \ce{ClCO3} as a function of height and pressure. Droplet chemistry is not included. The abundances of the chlorine radicals are greatly diminished in comparison to Figure \ref{fig:radicals_cloud_chem}. \label{fig:radicals_no_cloud_chem}}
\end{figure*}

\clearpage

\bibliography{bulk}{}
\bibliographystyle{aasjournal}

\bibliographystyle{aasjournal}

{}

\end{document}